\documentclass[usenatbib]{mnras}

\usepackage{graphicx}	% Including figure files
\usepackage{amsmath}	% Advanced maths commands
\usepackage{amssymb}	% Extra maths symbols
\usepackage{multicol}        % Multi column entries in tables
\usepackage{bm}		% Bold maths symbols, including upright Greek
\usepackage{pdflscape}	% Landscape pages
\usepackage{xcolor}     % Colors
\usepackage{orcidlink}  % orcid ids after names
\newcommand{\msun}{M_{\odot}}

\newcommand{\mpch}{$\mathrm{Mpc}\,h^{-1}$}
\newcommand{\gpc}{$\mathrm{Gpc}$}
\newcommand{\gpch}{$\mathrm{Gpc}\,h^{-1}$}

\newcommand{\disperse}{$\mathrm{DisPerSE \ }$}
\usepackage[T1]{fontenc}
\usepackage{ae,aecompl}

\usepackage{txfonts}

\title[Enabling DisPerSE at Gigaparsec Scales]{
Enabling Cosmic Web Analysis at Gigaparsec Scales: A Multi Block Approach for DisPerSE
}

\author[Singh et al.]{
Ankit Singh\orcidlink{0000-0001-5427-4515}$^{1}$\thanks{Contact e-mail: \href{mailto:ankit.singh@nottingham.ac.uk}{ankit.singh@nottingham.ac.uk}}, Frazer Pearce\orcidlink{0000-0002-2383-9250}$^{1}$, Meghan Gray\orcidlink{0000-0002-6301-5870}$^{1}$, 
Gustavo Yepes\orcidlink{0000-0001-5031-7936}$^{2,3}$,
\\
$^{1}$School of Physics and Astronomy, University of Nottingham, University Park, Nottingham NG7 2RD, United Kingdom\\
$^{2}$Departamento de Física Teórica M-8, Universidad Autónoma de Madrid, Cantoblanco, 28049, Madrid, Spain\\
$^{3}$Centro de Investigaciín Avanzada en Física Fundamental (CIAFF), Facultad de Ciencias, Universidad Autónoma de Madrid, Cantoblanco, 28049, Madrid, Spain
}

\date{Accepted XXX. Received YYY; in original form ZZZ}

\pubyear{2026}

\begin{document}
\label{firstpage}
\pagerange{\pageref{firstpage} \pageref{lastpage}}
\maketitle

\begin{abstract}
Cosmic filaments are the longest structures in the Universe and the dominant element of the cosmic web, channelling matter onto clusters and shaping the environments in which galaxies form and evolve. Accurate reconstructions of this network across gigaparsec volumes are increasingly important for cosmology and galaxy evolution. However, the most commonly used topological filament finder, \disperse(Discrete Persistent Structures Extractor), faces a memory bottleneck: it requires a Delaunay tessellation of the full input point set, preventing application to large simulations. Naively splitting the volume fails, as different sub-volumes yield inconsistent tessellations and filament networks. We present a frozen-core method that overcomes this bottleneck while preserving the global topology. The volume is decomposed into overlapping blocks whose tessellations are filtered by a circumsphere criterion retaining only globally valid tetrahedra; a post-processing pipeline merges the tiled outputs through core filtering, deduplication, and boundary stitching. Validation against a monolithic reference on a $300\,h^{-1}\,\mathrm{Mpc}$ MDPL2 subvolume shows 99.6 per cent total length recovery, 100 per cent recovery of density maxima and minima, and 94.7 per cent individual filament matching (the ${\sim}$5 per cent of unmatched filaments are predominantly short, low-significance structures). We apply the method to the full $(1\,h^{-1}\,\mathrm{Gpc})^3$ MDPL2 box (92 million haloes), producing a gigaparsec-scale filament catalogue. As a first application, we measure the connectivity ($\kappa$) for 22{,}900 haloes spanning $M_{200\mathrm{c}} = 10^{12}$--$10^{15.5}\,h^{-1}\,\mathrm{M}_\odot$, finding a power-law mass--connectivity relation that extends from group to cluster scales, providing the first confirmation in an $N$-body halo catalogue that the theoretically predicted scaling holds across three decades in halo mass.
\end{abstract}


%%%%%%%%%%%%%%%%%%%%%%%%%%%%%%%%%%%%%%%%%%%%%%%%%%

%%%%%%%%%%%%%%%%% BODY OF PAPER %%%%%%%%%%%%%%%%%%

\section{Introduction}
\label{sec:intro}

The matter distribution in the Universe on scales of megaparsecs is arranged in an intricate pattern of interconnected structures known as the cosmic web. Early observations revealed this structure of clusters, filaments, sheets, and voids \citep{deLapperant1986, Geller1989}. This network was anticipated theoretically using anisotropic gravitational collapse of primordial density perturbations \citep{Zeldovich1970}. \citet{Bond1996}, building on the peak statistics of \citet{Bardeen1986}, showed that the filamentary pattern is imprinted in the initial conditions of the density field: the tidal field between rare density peaks produces coherent bridges of overdensity that develop into the filaments observed in $N$ body simulations and galaxy surveys. The cosmic web framework has since become the standard paradigm for understanding the large-scale structure of the Universe \citep[for reviews, see][]{Cautun2014,Libeskind2018,Tojeiro2025}.

Large-scale filaments are the dominant structural element of the cosmic web by mass. Using the NEXUS+ multiscale morphological classification, \citet{Cautun2014} found that filaments contain approximately 50 per cent of the total mass at $z = 0$, compared to ${\sim}$11 per cent in clusters, ${\sim}$24 per cent in walls, and ${\sim}$15 per cent in voids. The Multiscale Morphology Filter yields a similar filamentary mass fraction of ${\sim}$39 per cent \citep{2010Aragon}. Hydrodynamical simulations predict that filaments also host a substantial fraction of the baryonic budget in the form of warm-hot intergalactic medium (WHIM) at $10^5$--$10^7$\,K \citep{CenOstriker1999, Dave2001}, with the remaining baryons locked in a cooler condensed phase at $T < 10^5$\,K associated with stars and the interstellar medium within galaxies \citep{Dave2001}. Filaments serve as the conduits through which matter flows from underdense regions toward the massive nodes of the cosmic web, channelling gas accretion onto galaxies and galaxy clusters.

The cosmic large-scale environment actively shapes the properties of galaxies residing in it. The effects discussed below operate on scales from a few hundred kpc to a few Mpc from the filament spine; the large-scale skeletal filaments identified in this work (typically 5--10\,\mpch\ long, traced by the halo distribution) represent the backbone of the cosmic web within which these smaller-scale processes occur. It is well established that galaxy properties depend on local density, the morphology--density relation \citep{Dressler1980} and the separability of mass and environmental quenching \citep{Peng2010}. Gas accretion onto galaxies proceeds preferentially along filaments, through cold-mode accretion flows that penetrate the virial radius without shock heating \citep{Keres2005, Dekel2009}. Galaxies closer to filament spines are observed to be statistically redder, more massive, and less star-forming than those further away, both in observations \citep{2017Kuutma, Chen2017, Malavasi2017, 2016Alpaslan, Darvish2015} and in simulations \citep{Kraljic2018, 2020Singh, Hasan2023}. The fraction of early-type morphologies increases toward the filament axis \citep{2017Kuutma, OKane2024}, and galaxy shapes preferentially align along filament spines \citep{EuclidLaigle2025}. Gas phase metallicities are elevated near cosmic web nodes and filaments \citep{Donnan2022, Rowntree2024, Rowntree2025}. The gas content of filament galaxies shows a mass-dependent behaviour: massive galaxies ($\log M_\ast/\msun > 11$) appear to accrete gas from the large-scale filament environment \citep{Odekon2018, Vulcani2019}, while lower-mass galaxies lose their gas reservoirs as they enter filaments \citep{Odekon2018, Hasan2024, Bulichi2023}. Whether these trends are driven purely by local density or require a distinct filament scale mechanism remains debated \citep{Song2021, OKane2024}. In the cluster infall regions, galaxies may undergo pre-processing within filaments and the groups embedded in them before entering the cluster environment \citep{Sarron2019, Donnan2022}. Understanding the geometry of the filamentary network around clusters, including the number of filaments connected to each cluster, known as the connectivity ($\kappa$), is therefore important for interpreting the accretion history and mass assembly of clusters, as the number and orientation of connected filaments sets the geometry of matter infall \citep{DarraghFord2019, Codis2018, Santoni2024}. Quantifying these connections requires a precise extraction of the filamentary network, a task that is complicated by the lack of a universal filament definition and the diversity of tracers and scales used across the literature.

Extracting the filamentary network from these data requires specialised algorithms, but the field currently lacks a standardised definition of what constitutes a filament: methods differ fundamentally in how they partition the cosmic web, and catalogues produced by different approaches are not directly comparable \citep{Libeskind2018}. Studies also employ diverse tracers -- dark matter particles and gas in hydrodynamical simulations, halo catalogues in $N$-body simulations, and photometric or spectroscopic galaxy catalogues in observations -- and probe filaments across a wide range of scales, from individual inter-cluster bridges of a few $h^{-1}$\,Mpc to the large-scale skeletal structure of the web on scales of tens of $h^{-1}$\,Mpc, which are accessible only in large simulation volumes. More than a dozen distinct algorithms have been proposed \citep[for a comprehensive comparison, see][]{Libeskind2018}, ranging from geometric approaches such as the Multiscale Morphology Filter \citep{2010Aragon} and the Bisous model \citep{Tempel2014}, to density field classification schemes such as NEXUS/NEXUS+ \citep{Cautun2013, Cautun2014}, to topological methods based on Morse theory. Among these, \disperse \citep[Discrete Persistent Structures Extractor;][]{2011aSousbie} stands out for its topological rigour and applicability to discrete point distributions such as halo or galaxy catalogues: it constructs the full Morse-Smale complex of the density field and uses persistent homology \citep{Edelsbrunner2008} to identify structures at all significance levels with a single user-specified threshold $n_\sigma$ (the significance level, in units of the root-mean-square density fluctuation expected from Poisson noise, above which topological features are retained; see Appendix~\ref{sec:parameter_sensitivity}), making it straightforward to compare results across different tracer densities and survey geometries. \disperse has been adopted widely for filament detection in both simulations \citep{2020Singh,Galarraga2021, Galarraga2022, GalarragaEspinosa2024} and observations \citep{Laigle2018, Mahajan2018, Malavasi2020}. In observations, the identification of filaments faces additional challenges beyond those present in simulations, most notably redshift-space distortions that elongate structures along the line of sight and can spuriously connect or fragment filaments \citep{Kuchner2021}.

In brief, \disperse identifies filaments as density ridges in a point distribution by: (i)~constructing a Delaunay tessellation (a space-filling mesh of tetrahedra) from the input points; (ii)~estimating the local density at each point from the Voronoi cell volumes; (iii)~locating critical points -- peaks, saddles, and minima -- of this density field; and (iv)~tracing the gradient flow lines that connect these critical points into the Morse--Smale complex, whose one-dimensional arcs are the filaments. A single user-specified threshold, $n_\sigma$, controls the significance level above which features are retained. A formal description of each step is given in Section~\ref{sec:disperse_bottleneck}.

Despite the success of \disperse, the monolithic three-dimensional Delaunay tessellation that it requires as input creates a computational bottleneck when applied to large halo catalogues: the tessellation must load the entire point set into RAM, with requirements scaling as ${\sim}5$--$10$\,GB per million input particles. For the MDPL2 simulation \citep{Klypin2016}, which contains 92 million haloes in a $(1\,h^{-1}\,\mathrm{Gpc})^3$ box (after the 20-particle mass cut adopted in this work), the Delaunay tessellation alone would require ${\sim}460$--$920$\,GB of RAM, far exceeding the capacity of standard compute nodes. \disperse has been applied to comparable or larger simulations -- \citet{GalarragaEspinosa2024} ran it on the MillenniumTNG galaxy catalogue (${\sim}3 \times 10^6$ subhaloes with $M_\ast > 10^9\,\msun$) in a $(500\,h^{-1}\,\mathrm{Mpc})^3$ box, and \citet{Kumar2026} applied it to galaxies ($M_\ast \geq 10^9\,\msun$) in the $(1\,\mathrm{cGpc})^3$ FLAMINGO simulation -- but these analyses use stellar-mass-selected galaxy catalogues containing far fewer tracers than the full halo catalogue. No study has yet tessellated a dark matter halo catalogue of ${\sim}10^8$ objects monolithically (i.e.\ processing the entire volume as a single unit). The current generation of cosmological simulations, including FLAMINGO \citep{Schaye2023} and the forthcoming Euclid Flagship \citep{EuclidFlagship2024}, routinely produce halo catalogues of $10^8$--$10^9$ objects for which monolithic \disperse runs become increasingly impractical. A naive solution of splitting the volume into independent sub-volumes and running \disperse separately on each is not viable without careful control: the Delaunay tessellation and hence the persistence pairs that define the filament topology depend on the global point distribution, so different sub-volumes produce different network topologies even in their shared regions. Simply concatenating independently detected sub-volume networks therefore lacks internal consistency and cannot recover the filaments that a full-box run would find; as we demonstrate in Section~\ref{sec:disperse_bottleneck}, the discrepancy is systematic and cannot be corrected by adjusting the persistence threshold.

In this paper, we present a \emph{frozen-core} method -- so called because the interior (core) region of each tile has its Delaunay tessellation `frozen' by the surrounding padding, ensuring it is identical to the global result -- that enables \disperse to analyse Gpc-scale halo catalogues while preserving the topological integrity of the filament network. We validate the method against a monolithic reference on a $300\,h^{-1}\,\mathrm{Mpc}$ subvolume of MDPL2 and apply it to the full 1\,\gpc\ MDPL2 box, producing the first gigaparsec-scale filament catalogue obtained with topologically rigorous methods. This opens access to filaments on scales of tens of Mpc and to the statistical power of a large cosmological volume unavailable in smaller runs. As a first scientific application, we measure the connectivity $\kappa$ for 22{,}900 haloes spanning three decades in mass from galaxy groups to the most massive clusters in MDPL2, including the 324 parent haloes of The Three Hundred project \citep{Cui2018}, and compare with theoretical predictions and previous measurements from gas-based analyses.

The paper is organised as follows. Section~\ref{sec:mdpl2} describes the MDPL2 simulation and input halo catalogue. Section~\ref{sec:disperse_bottleneck} summarises the \disperse algorithm and identifies the two memory bottlenecks -- the Delaunay tessellation and the Morse--Smale extraction -- and Section~\ref{sec:naive_tiling} explains why naively tiling the tessellation is insufficient. Section~\ref{sec:padding} determines the padding width from the tracer circumradius distribution and validates it with a controlled boundary-effect experiment. Section~\ref{sec:frozen_core} details the three-stage frozen-core pipeline -- core filtering, spatial deduplication, and boundary stitching -- and validates it against the monolithic reference. Section~\ref{sec:results} presents the 1\,\gpc\ filament catalogue, its sensitivity to the persistence threshold, and the cluster connectivity analysis. Section~\ref{sec:discussion} discusses the implications and limitations, and Section~\ref{sec:conclusions} summarises our conclusions. Appendix~\ref{sec:parameter_sensitivity} presents parameter sensitivity tests.

\section{Method}
\label{sec:method}

We begin by describing the MDPL2 simulation and input halo catalogue (\S\,\ref{sec:mdpl2}), then present the \disperse algorithm and its memory bottleneck (\S\,\ref{sec:disperse_bottleneck}). The frozen-core method proceeds in three stages. First, we decompose the simulation volume into overlapping spatial blocks whose size is chosen so that the Delaunay tessellation of each block fits in available RAM (\S\,\ref{sec:padding}). Each block consists of an inner \emph{core} region from which filaments will be extracted, surrounded by a \emph{padding} layer of width $P$ that provides the spatial context needed for a correct tessellation. Second, we compute the tessellation independently on each padded block, then apply a \emph{circumsphere}-based filter that discards tetrahedra whose circumspheres extend beyond the padded region -- these are not guaranteed to belong to the global tessellation. The tetrahedra that survive this filter constitute the \emph{frozen core}: their tessellation is identical to what a monolithic run would produce, because all points that could affect them are present within the padded block. Third, we run \disperse on the merged, filtered tessellation and apply a filament post-processing pipeline (\S\,\ref{sec:frozen_core}) that removes duplicates arising from tile overlaps and stitches fragments at block boundaries.

\subsection{The MDPL2 Simulation}
\label{sec:mdpl2}

We use the MultiDark Planck 2 (MDPL2) $N$-body simulation \citep{Klypin2016}, one of the MultiDark suite of cosmological simulations publicly available through the CosmoSim database \citep{Riebe2013}. MDPL2 follows the evolution of $3840^3$ dark matter particles in a periodic box of side length $L = 1\,h^{-1}\,\mathrm{Gpc}$, from an initial redshift of $z = 120$ to $z = 0$. The particle mass is $m_p = 1.51 \times 10^9\,h^{-1}\,\msun$, and the gravitational force resolution ranges from 5 to 13\,$h^{-1}\,\mathrm{kpc}$ (physical) depending on redshift. The simulation adopts a flat $\Lambda$CDM cosmology consistent with \citet{Planck2014}: $\Omega_\mathrm{m} = 0.307$, $\Omega_\Lambda = 0.693$, $\Omega_\mathrm{b} = 0.048$, $h = 0.678$, $\sigma_8 = 0.823$, and $n_\mathrm{s} = 0.96$. Dark matter haloes are identified using the \textsc{Rockstar} phase space halo finder \citep{Behroozi2013}. In this work, we use the $z = 0$ Rockstar catalogue, which contains 127,388,160 haloes spanning a virial mass range of $3 \times 10^9$ to $3.5 \times 10^{15}\,h^{-1}\,\msun$. We apply a 20-particle minimum mass cut ($M_{\rm vir} \geq 3.02 \times 10^{10}\,h^{-1}\,\msun$), retaining 92,282,235 haloes (72.4 per cent by number, 98.0 per cent by mass). Haloes resolved with fewer than 20 dark matter particles have poorly constrained positions and masses \citep{Behroozi2013}, and their inclusion introduces noise into the Delaunay tessellation without improving the density field at the scales relevant for filament detection ($\gtrsim 1$\,\mpch). The mass cut increases the mean inter-halo separation from ${\sim}2.0$ to ${\sim}2.2$\,\mpch, corresponding to a number density of ${\sim}0.092$ haloes per $(\mathrm{Mpc}\,h^{-1})^3$. These halo positions serve as the input point set for the Delaunay tessellation and subsequent filament detection.

\subsection{The \texorpdfstring{\disperse}{DisPerSE} algorithm}
\label{sec:disperse_bottleneck}

We summarise the \disperse algorithm to define the key technical terms used throughout this paper and to identify the specific step that creates the memory bottleneck. \disperse \citep{2011aSousbie, 2011bSousbie} constructs a discrete Morse--Smale complex from the three-dimensional input particle distribution through the following sequence:

\begin{enumerate}

    \item \textbf{Delaunay tessellation.} The input point set (particle positions or halo centres) is tessellated into tetrahedra in three dimensions, producing a \emph{simplicial complex}: a collection of simplices -- vertices (0D), edges (1D), triangular faces (2D), and tetrahedra (3D) -- that fit together so that any two simplices are either disjoint or share a complete lower-dimensional face. This complex encodes the full spatial connectivity of the data.
    \item \textbf{Density estimation.} A density value is assigned to each vertex using the Delaunay Tessellation Field Estimator \citep[DTFE;][]{Schaap2000}. For each vertex, the DTFE computes the volume of its Voronoi cell (the region of space closer to that vertex than to any other) and estimates the local density as the total mass associated with the vertex divided by the cell volume. This yields a continuous, volume-weighted density field sampled at the input point positions.
    \item \textbf{Discrete gradient.} A discrete gradient field is computed on the simplicial complex by pairing simplices of adjacent dimension -- that is, simplices differing by one in dimensionality (e.g.\ an edge with one of its bounding vertices, or a triangle with one of its bounding edges) -- to define flow directions from low- to high-density regions.
    \item \textbf{Critical points and Morse--Smale complex.} Unpaired simplices are identified as critical points: minima (type 0), 1-saddles (type 1), 2-saddles (type 2), and maxima (type 3) in three dimensions. The ascending and descending manifolds connecting these critical points define the Morse--Smale complex. In the \disperse framework, a \emph{filament} is defined as the one-dimensional arc of the Morse--Smale complex connecting a density maximum (type-3 critical point) to a 2-saddle (type-2 critical point) via the density gradient. Each filament is represented as an ordered sequence of sampling points tracing this arc through the simplicial complex.
    \item \textbf{Persistence computation.} Persistence pairs are computed from the \emph{topological boundary matrix} of the simplicial complex -- the matrix that encodes which lower-dimensional simplices form the boundary of each higher-dimensional one (e.g.\ which three edges bound a given triangle). Each persistence pair associates a critical point that creates a topological feature with the critical point that destroys it; the persistence of a pair is the density contrast between the two critical points, providing a scale-independent measure of feature significance: high-persistence filaments survive a wide range of $n_\sigma$ thresholds, while low-persistence ones are suppressed at higher values.
    \item \textbf{Simplification.} Persistence pairs below a user-specified significance threshold $n_\sigma$ are cancelled, removing low-contrast features and retaining only structures with density contrast exceeding $n_\sigma$ times the expected root-mean-square density fluctuation due to Poisson noise in the discrete point set.

\end{enumerate}

The critical point for this work is that persistence pairs (step~5) depend on the topological boundary matrix, which is determined by the simplicial complex constructed in step~1. Changing the simulation volume changes the Delaunay tessellation, which changes the network topology, which changes the persistence pairs even if the density values at shared vertices are identical. This is why, as we will show in Section~\ref{sec:frozen_core}, running \disperse on individual smaller sub volumes produces results that are not consistent with extracting the same subsets from the network constructed from the full box.

The computational bottleneck in \disperse arises from two stages that each require the full dataset to reside in RAM simultaneously. The first is the three-dimensional Delaunay tessellation (step~1), which must load the entire point set and build the growing tessellation incrementally. The second is the Morse--Smale extraction (MSE, steps~3--5), which loads the complete simplicial complex to compute the discrete gradient and persistence pairs. Both stages are single-threaded in the current \disperse implementation, so their RAM requirements cannot be reduced by adding more CPU cores. We quantify each below.

\subsubsection{Delaunay tessellation (RAM)} For a three-dimensional point distribution that is close to Poissonian on large scales, the number of Delaunay tetrahedra scales linearly with the number of points, with
  \[ N_{\mathrm{tet}} \simeq (6\text{-}7)\,N \]
  \citep[e.g.][]{1994vandeWeygaert,Okabe2000}. Adopting a conservative
  factor $N_{\mathrm{tet}} = 6N$, the mass-cut MDPL2 halo catalogue produces
  \[ N_{\mathrm{tet}} \simeq 6 \times 9.2 \times 10^{7} = 5.5 \times 10^{8}\]
  tetrahedra. Each tetrahedron must store vertex connectivity and adjacency information. In practical implementations this corresponds to a memory footprint of at least ${\sim}\,64$~bytes per tetrahedron once indexing, alignment, and bookkeeping overheads are included. The storage of the simplicial complex alone is therefore
  \[ M_{\mathrm{DT}} \simeq N_{\mathrm{tet}} \times 64~\mathrm{bytes} \simeq 3.5 \times 10^{10}~\mathrm{bytes} \approx 35~\mathrm{GB}.\]
  This is a lower bound: the incremental insertion algorithm also requires working RAM for point location, conflict zone management, and temporary structures. Empirically, modern Delaunay codes consume ${\sim}\,5$--$10$~GB of RAM per million input particles, placing the full Delaunay construction at ${\sim}\,460$--$920$~GB for the mass-cut MDPL2 catalogue.

\begin{figure}
    \centering
    \includegraphics[width=\columnwidth]{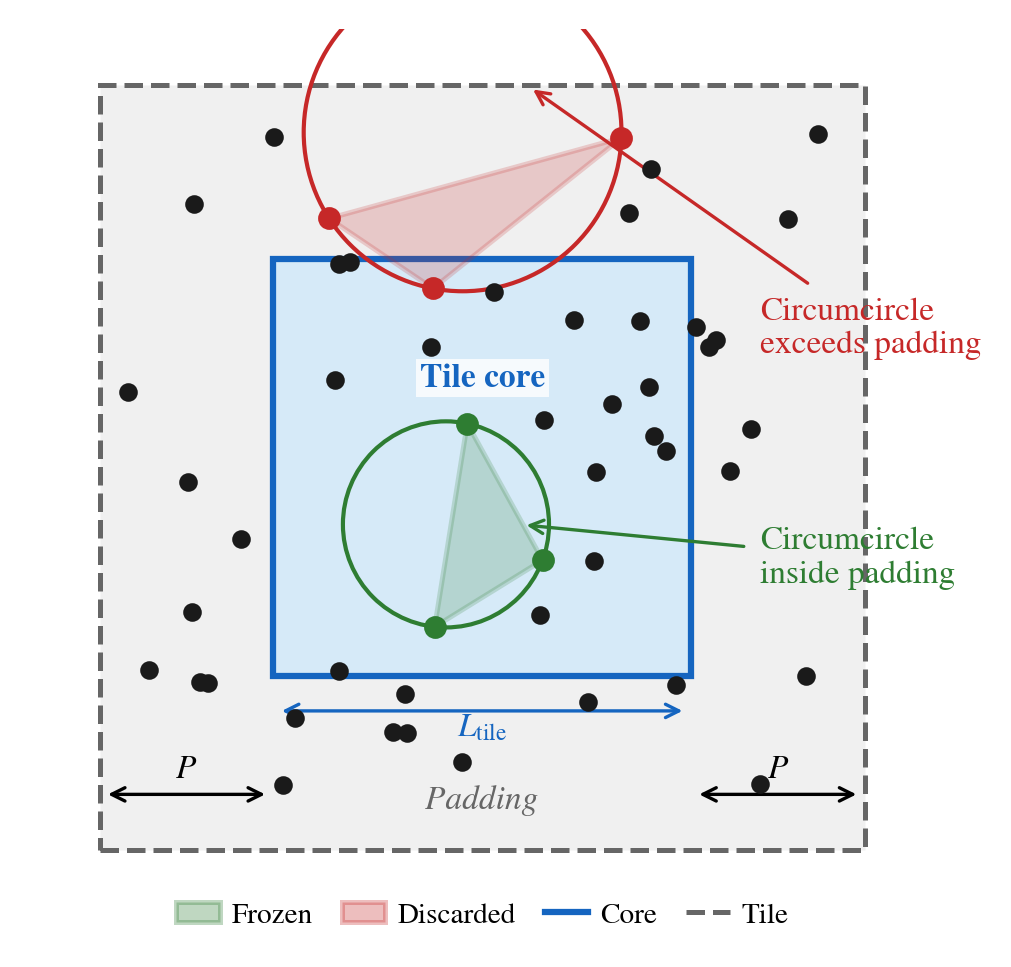}
    \caption{Schematic illustration of the frozen-core concept in 2D. The inner \emph{core} region (blue, solid boundary) is surrounded by a \emph{padding} layer of width $P$ (grey, dashed boundary). Delaunay triangles (the 2D analogue of tetrahedra) in the core are classified by their circumcircles: a triangle whose circumcircle fits entirely within the padded tile (green) is \emph{frozen} -- guaranteed to match the global tessellation -- and retained. A triangle whose circumcircle extends beyond the padding (red) may be invalid and is discarded.}
    \label{fig:schematic}
\end{figure}

\begin{figure*}
    \centering
    \includegraphics[width=0.9\textwidth]{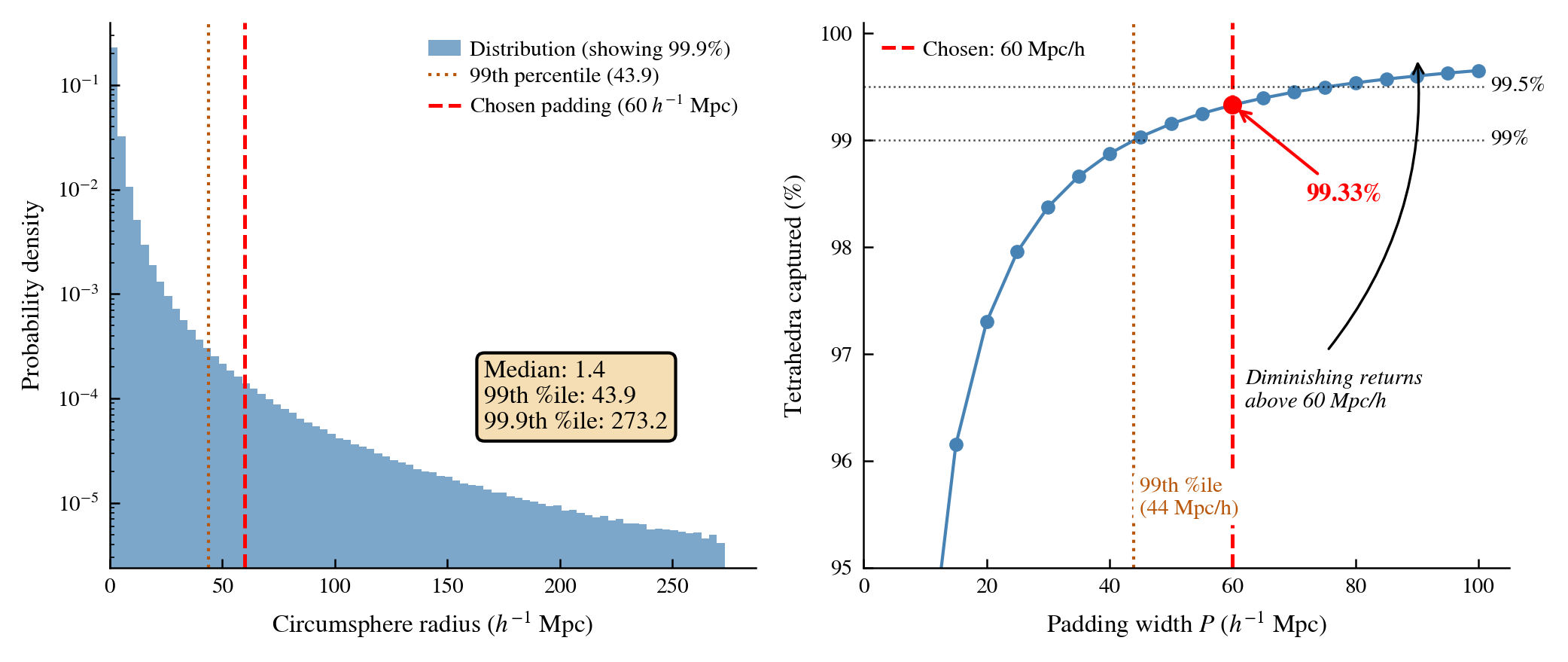}
    \caption{\textbf{Circumsphere radius distribution and padding selection} for Delaunay tetrahedra in the padded validation volume (300\,\mpch\ MDPL2 subvolume, 20-particle mass cut, 6.8 million haloes, 44.7 million tetrahedra). \textbf{Left:} Probability density of circumradii (truncated at the 99.9th percentile). The distribution is highly skewed (median 1.4\,\mpch, 99th percentile 43.9\,\mpch) due to large tetrahedra in voids. The chosen 60\,\mpch\ padding (red dashed) exceeds the 99th percentile by $1.4\times$. \textbf{Right:} Cumulative fraction of tetrahedra captured versus padding width. The 60\,\mpch\ padding captures 99.33 per cent, with diminishing returns beyond this point (equation~\ref{eq:volume_overhead}).}
    \label{fig:circumradius_dist}
\end{figure*}

\subsubsection{Morse--Smale extraction (RAM)} After tessellation, the MSE step loads the full simplicial complex into RAM and computes a discrete gradient on every simplex. The total number of simplices in a 3D Delaunay complex scales as
  \[ N_{\mathrm{simp}} = N_{\mathrm{vert}} + N_{\mathrm{edge}} + N_{\mathrm{face}} + N_{\mathrm{tet}} \simeq N + 7N + 12N + 6N = 26\,N, \]
  where the edge and face counts follow from the Euler relation for Delaunay complexes \citep{Okabe2000}. Each simplex requires a gradient pairing entry and associated bookkeeping (${\sim}\,8$--$16$~bytes), adding ${\sim}\,26N \times 12~\mathrm{bytes} \approx 40$~GB for the MDPL2 catalogue. Including the loaded complex and working buffers for persistence pair computation, the MSE peak memory is comparable to the Delaunay storage, reaching ${\sim}\,80$--$100$~GB.

  \paragraph{Total RAM.} The combined peak RAM is dominated by the Delaunay construction step, which requires the full point set, the growing tessellation, and algorithmic working space simultaneously in memory. The MSE step is less demanding because it operates on the already-constructed complex without the overhead of incremental insertion. Together, these costs make the standard monolithic \disperse\ pipeline prohibitive for $\mathcal{O}(10^{8})$ halo catalogues and completely infeasible for particle-level analyses of Gpc-scale simulations. Since both stages are single-threaded, the frozen-core method's key advantage is that each tile can be processed independently on a separate compute node, making the tessellation stage embarrassingly parallel while keeping peak RAM per node within standard limits (at most ${\sim}10$\,GB for our tile configuration).

\subsection{The memory bottleneck and limitations of naive tiling}
\label{sec:naive_tiling}

A natural first approach is to compute the Delaunay tessellation independently on overlapping tiles and merge the results before running the MSE step. While this reduces peak memory during tessellation, it introduces two problems. First, overlapping tiles produce redundant tetrahedra in the padding regions: in our 1\,\gpc\ run, the raw tile outputs contain a ${\sim}\,2.6\times$ overcount of tetrahedra relative to those retained after circumsphere-based deduplication. Without filtering, the MSE step must load this inflated complex, negating the memory savings of tiling. Second, the duplicated simplices violate the requirement that the input form a valid simplicial complex (each simplex appearing exactly once), causing the discrete Morse gradient computation to fail.

Our frozen core method (\S\,\ref{sec:frozen_core}) resolves both issues: it removes redundant tetrahedra, restoring a valid simplicial complex, and reduces the merged output to within ${\sim}\,1$ per cent of the monolithic tetrahedron count. The MSE step then operates on a complex of the same size as the full box case, but constructed with bounded peak memory.

\begin{figure*}
    \centering
    \includegraphics[width=\textwidth]{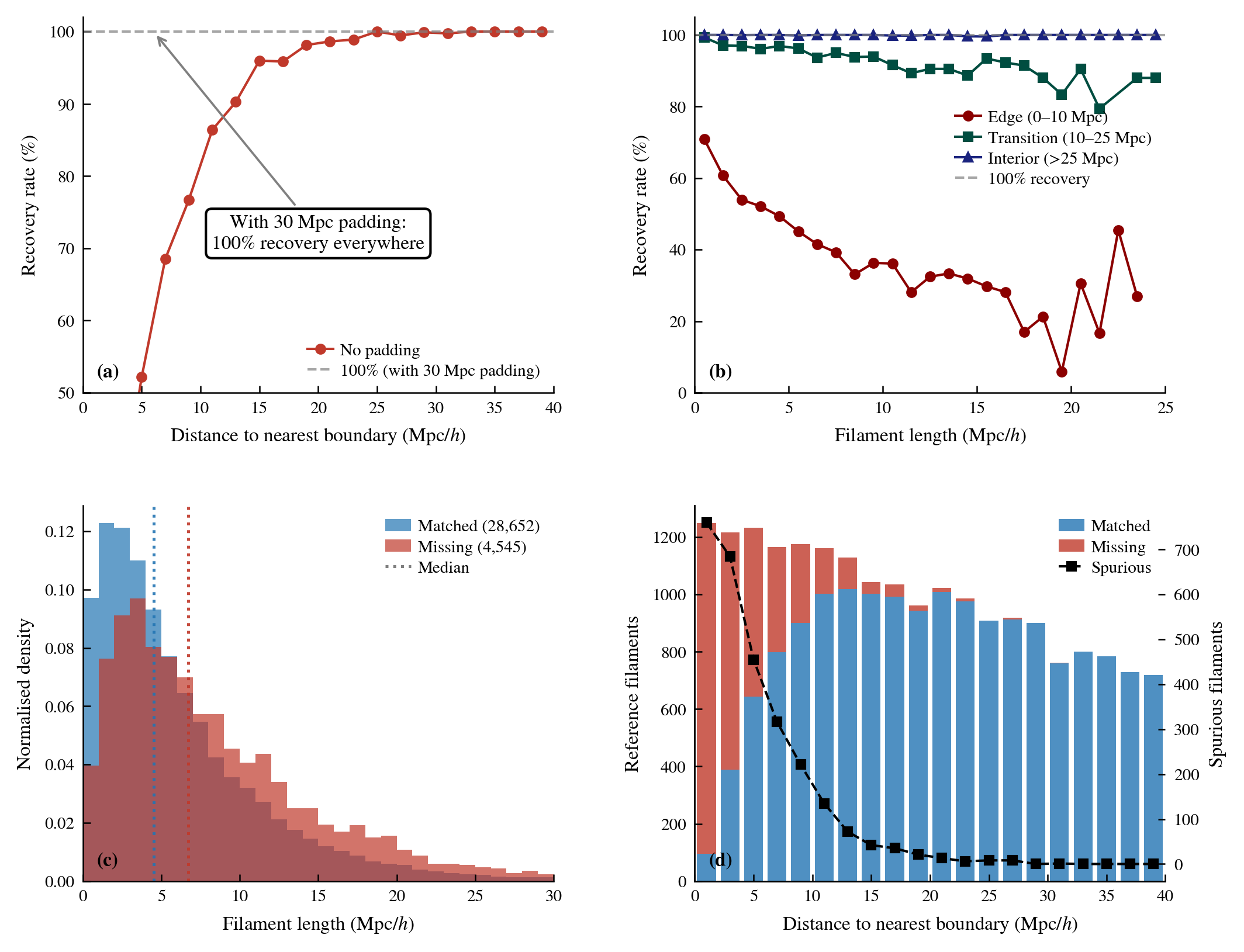}
    \caption{Boundary effects on filament detection in a 300\,\mpch\ sub-volume without padding, compared to the fully-padded 60\,\mpch\ reference, which reproduces the monolithic result ($n_\sigma = 6.5$). Filaments are matched by centroid proximity ($<0.5$\,\mpch; centroid = mean position of all sampling points). Even half the fiducial padding, 30\,\mpch, already achieves 100 per cent recovery everywhere (annotated in panel~a), so the 60\,\mpch\ reference carries ample margin.
    \textbf{(a)}~Recovery rate versus boundary distance: drops sharply near the edge; the y-axis is truncated at 50 per cent to show the trend clearly, but recovery falls to ${\sim}$20 per cent within 5\,\mpch\ of the edge (see panel~d for counts).
    \textbf{(b)}~Recovery versus filament length by spatial zone. Edge filaments ($<10$\,\mpch\ from boundary) with $L>20$\,\mpch\ fall below 10 per cent recovery; interior filaments ($>25$\,\mpch) reach ${\sim}$100 per cent at all lengths.
    \textbf{(c)}~Normalised length distributions of matched (blue) and missing (red) reference filaments; dotted lines mark medians. Missing filaments are systematically longer (median $6.7$ vs $4.5$\,\mpch), since longer filaments are more likely to span the boundary, where the absent context corrupts the tessellation.
    \textbf{(d)}~Matched and missing reference filaments by boundary distance (stacked bars) and spurious no-padding filaments (right axis, black). Both the missing and the spurious populations are confined within ${\sim}10$\,\mpch\ of the boundary.}
    \label{fig:boundary_recovery}
\end{figure*}

\subsection{Padding Determination}
\label{sec:padding}

The first methodological choice in the frozen-core method is the padding width $P$ surrounding each tile's core region. $P$ must be large enough so that no point lying outside the padded volume could fall inside the circumsphere of any core tetrahedron, which would violate the empty-circumsphere property that defines a valid Delaunay tetrahedron (i.e.\ no input point other than its four vertices lies inside its circumsphere). The circumsphere of a tetrahedron is the unique sphere passing through its four vertices; its radius (the circumradius) is the distance from the circumcentre to any vertex. A tetrahedron is therefore guaranteed to belong to the global Delaunay tessellation if its circumsphere fits entirely within the padded region (Figure~\ref{fig:schematic}). We determine $P$ empirically from the circumradius distribution.

\subsubsection{Circumradius Distribution Analysis}

We analyse the circumsphere radius distribution from the Delaunay tessellation of the padded validation volume (300\,\mpch\ MDPL2 subvolume core with 60\,\mpch\ padding on each side, using the mass cut of Section~\ref{sec:mdpl2}), comprising 6.8 million haloes and 44.7 million tetrahedra (number density ${\sim}0.092$~haloes per $(\mathrm{Mpc}\,h^{-1})^{3}$, consistent with the full 1\,\gpc\ box). The circumsphere analysis uses the full padded tessellation rather than the 300\,\mpch\ core alone, because the padding region provides the surrounding context that determines the circumsphere radii of tetrahedra near the core boundary. The mean inter-halo separation in this volume is ${\sim}2.2$\,\mpch, matching that of the full 1\,\gpc\ run. Figure~\ref{fig:circumradius_dist} shows the distribution, and Table~\ref{tab:circumradius} summarises the statistics.

\begin{table}
    \centering
    \caption{Circumsphere radius distribution statistics from the full Delaunay tessellation of the padded validation volume (300\,\mpch\ MDPL2 subvolume core with 60\,\mpch\ padding; 20-particle mass cut, 6.8 million haloes, 44.7 million tetrahedra). The 60\,\mpch\ padding captures 99.3 per cent of all tetrahedra.}
    \label{tab:circumradius}
    \begin{tabular}{lc}
        \hline
        \textbf{Statistic} & \textbf{Value (\mpch)} \\
        \hline
        Median & 1.4 \\
        95th percentile & 12.1 \\
        99th percentile & 43.9 \\
        99.9th percentile & 273.2 \\
        \hline
        \textbf{Padding} & \textbf{Coverage (\%)} \\
        \hline
        40\,\mpch & 98.87 \\
        50\,\mpch & 99.15 \\
        60\,\mpch & 99.33 \\
        80\,\mpch & 99.54 \\
        \hline
    \end{tabular}
\end{table}

The distribution is highly skewed: the median circumradius is 1.4\,\mpch, while the 99th percentile is 43.9\,\mpch. The tail is long because large tetrahedra form preferentially in underdense void regions where the inter-halo spacing is large. We adopt a padding width of $P = 60$\,\mpch, which exceeds the 99th percentile by a factor of 1.4 and captures 99.33 per cent of all tetrahedra. The right panel of Figure~\ref{fig:circumradius_dist} shows the trade-off between padding width and tetrahedra captured: the curve exhibits clear diminishing returns above ${\sim}60$\,\mpch, with only 0.32 per cent additional coverage gained by increasing the padding from 60 to 100\,\mpch, at a volume overhead cost of $(220/160)^3 - 1 = 2.7\times$. This makes 60\,\mpch\ an appropriate choice that balances topology preservation against computational cost.

The padding introduces a volume overhead that must be considered when choosing the tile size. Each cubic tile has a core region of side length $L_{\rm tile}$ from which filaments are extracted. To ensure topology preservation, we surround this core with a padding layer of width $P$ on all six faces, so that the padded tile on which the Delaunay tessellation is actually computed has side length $L_{\rm tile} + 2P$. The volume of the padded tile is therefore $(L_{\rm tile} + 2P)^3$, and the volume overhead relative to the core is
\begin{equation}
\frac{V_{\rm padded}}{V_{\rm core}} = \frac{(L_{\rm tile} + 2P)^3}{L_{\rm tile}^3} = \left(1 + \frac{2P}{L_{\rm tile}}\right)^3.
\label{eq:volume_overhead}
\end{equation}
For our choice of $L_{\rm tile} = 100$\,\mpch\ and $P = 60$\,\mpch, this yields an overhead factor of $(1 + 1.2)^3 = 10.6$. The padded tile has side length $100 + 2 \times 60 = 220$\,\mpch, which contains ${\sim}1$ million haloes depending on the local density, and its Delaunay tessellation requires at most ${\sim}10$\,GB of RAM. We note that the padding width $P$ is determined entirely by the circumradius distribution of the tracer population and is therefore independent of the tile size $L_{\rm tile}$; the same $P = 60$\,\mpch\ applies to any tiling of the same halo catalogue, and only $L_{\rm tile}$ needs to be adjusted to match the available RAM per node.

% Tiling Validation Figures

\begin{figure*}
\centering
\includegraphics[width=\textwidth]{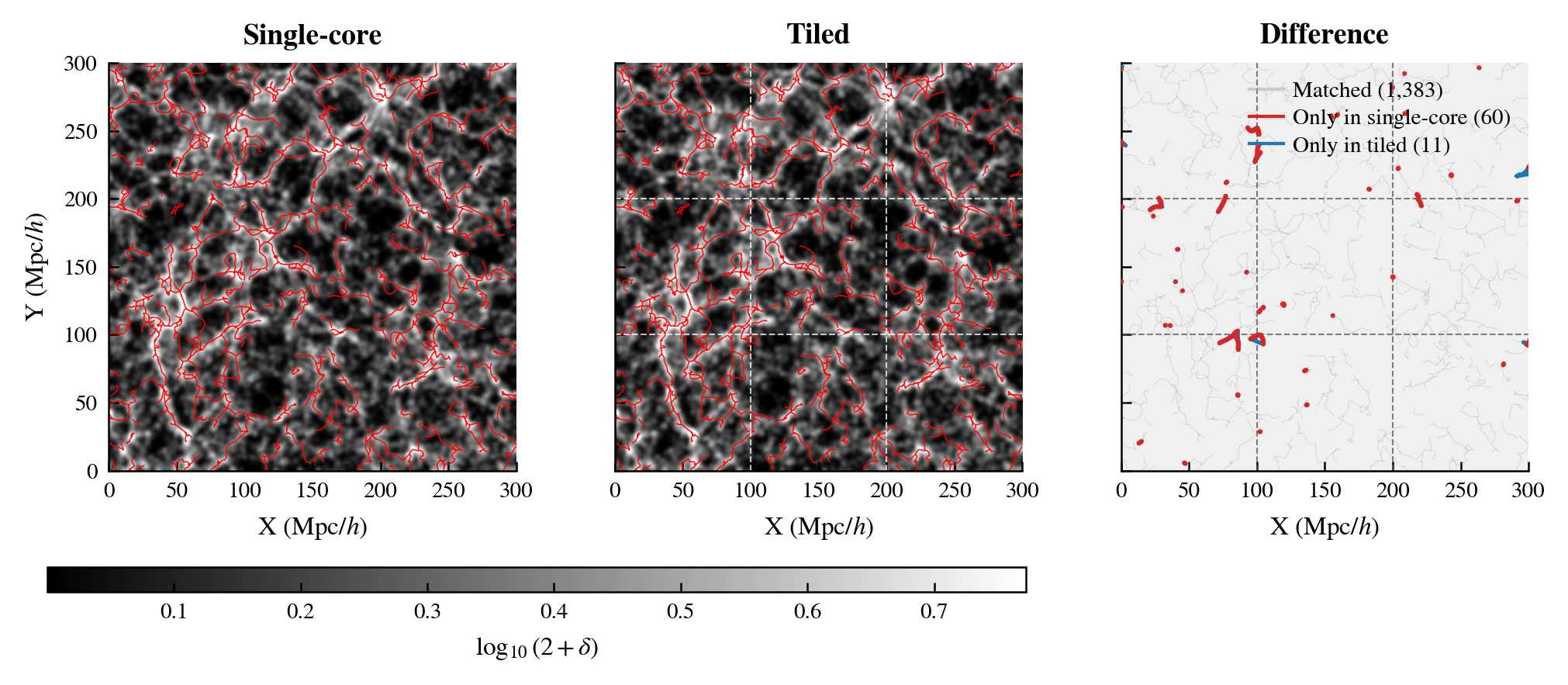}
\caption{Visual comparison of filament detection between single-core reference and tiled approach ($n_\sigma = 7.0$, smoothing~$= 8$, 20-particle mass cut) in a 10\,\mpch-thick slice at $z = 150$\,\mpch.
    \textbf{Left:} monolithic reference overlaid on the log-scaled halo density field ($n = 1{,}443$ filaments in slice).
    \textbf{Centre:} tiled approach ($n = 1{,}394$); dashed white lines mark tile boundaries at 100 and 200\,\mpch.
    \textbf{Right:} difference plot. Grey lines: 1{,}383 matched filaments. Red ($n = 60$): reference-only; blue ($n = 11$): tiled-only. The asymmetry confirms the pipeline is conservative, losing filaments without introducing false structure. Longer missing filaments cluster near tile boundaries, while shorter misses are distributed throughout the volume (see Fig.~\ref{fig:tiling_unmatched}b and Table~\ref{tab:tiling_validation}).}
\label{fig:tiling_visual}
\end{figure*}

\subsubsection{Empirical Validation: Boundary Effects Without Padding}

The theoretical circumradius argument motivates padding, but does not directly quantify the impact on detected filaments. We therefore perform a controlled experiment on a 300\,\mpch\ MDPL2 subvolume -- chosen so that the monolithic \disperse reference run, including 60\,\mpch\ padding on each side, requires at most ${\sim}55$\,GB of RAM, well within standard compute node capacities. We run \disperse on three extractions from the same region, with identical parameters ($n_\sigma = 6.5$, smooth~$= 8$), differing only in the amount of surrounding context:
\begin{enumerate}
    \item \textbf{Reference} ($300 + \mathrm{padding}$): 300\,\mpch\ MDPL2 subvolume core with 60\,\mpch\ padding on each side.
    \item \textbf{Partial padding} ($300 + \mathrm{half{-}padding}$): 300\,\mpch\ core with 30\,\mpch\ padding.
    \item \textbf{No padding} (300\,\mpch\ only): Core region only, no surrounding context.
\end{enumerate}

In all three cases, we run \texttt{delaunay\_3D} (the Delaunay tessellation tool of the \disperse toolkit) with its default open (non-periodic) boundary condition, as the padding itself provides the surrounding spatial context; no periodic or mirror boundary treatment is applied. We extract filaments from the shared 300\,\mpch\ inner region in all three cases and match them by the proximity of their centroids, defined as the mean position of all sampling points along each filament (tolerance $<0.5$\,\mpch). Figure~\ref{fig:boundary_recovery} presents a detailed analysis of the boundary effects.

Without padding, the recovery rate drops sharply near the boundary: only ${\sim}$20 per cent of filaments within 5\,\mpch\ of the edge are correctly recovered, and the boundary affected zone extends ${\sim}20$\,\mpch\ inward (panel~a). Overall, 13.7 per cent of filaments (4{,}548 out of 33{,}197) are lost or altered. With 30\,\mpch\ padding, recovery is 100 per cent at all distances, demonstrating that even moderate padding fully eliminates boundary artefacts. The 60\,\mpch\ padding used in our pipeline therefore provides ample margin.

The boundary effect is strongly length dependent (panel~b): in the edge zone (0--10\,\mpch\ from the boundary), recovery drops from ${\sim}$59 per cent for the shortest filaments ($L < 3$\,\mpch) to ${\sim}$13 per cent for filaments longer than 20\,\mpch. This is because longer filaments are more likely to extend across the boundary, where the missing context corrupts the Delaunay tessellation and alters the persistence pairs. The missing filaments are correspondingly shifted to longer lengths compared to the matched population (panel~c). In addition to losing genuine filaments, the boundary also creates spurious filament structures present in the no padding catalogue that have no counterpart in the reference (panel~d, black dashed line, right axis). These spurious filaments arise from artificial critical points at the tessellation boundary and, like the missing filaments, are concentrated within ${\sim}10$\,\mpch\ of the edge. Table~\ref{tab:boundary_effects} summarises the statistics.

\begin{table*}
\centering
\caption{Boundary effects on filament detection. Filament statistics for the inner 300\,\mpch\ core region from DisPerSE runs with varying amounts of surrounding context (padding). The 60\,\mpch\ padding run serves as the reference.}
\label{tab:boundary_effects}
\begin{tabular}{lrrr}
\hline
Metric & 60\,\mpch\ pad & 30\,\mpch\ pad & No padding \\
\hline
Filament count & 33,197 & 33,202 & 31,426 \\
Total length (\mpch) & 214,531 & 214,538 & 204,345 \\
Mean length (\mpch) & 6.46 & 6.46 & 6.50 \\
Median length (\mpch) & 4.76 & 4.76 & 4.74 \\
\hline
Recovery (overall) & 100\% & 100.0\% & 86.3\% \\
Recovery (0--10\,\mpch\ from edge) & 100\% & 99.9\% & 40.6\% \\
\hline
\end{tabular}
\end{table*}

\begin{figure*}
    \centering
    \includegraphics[width=\textwidth]{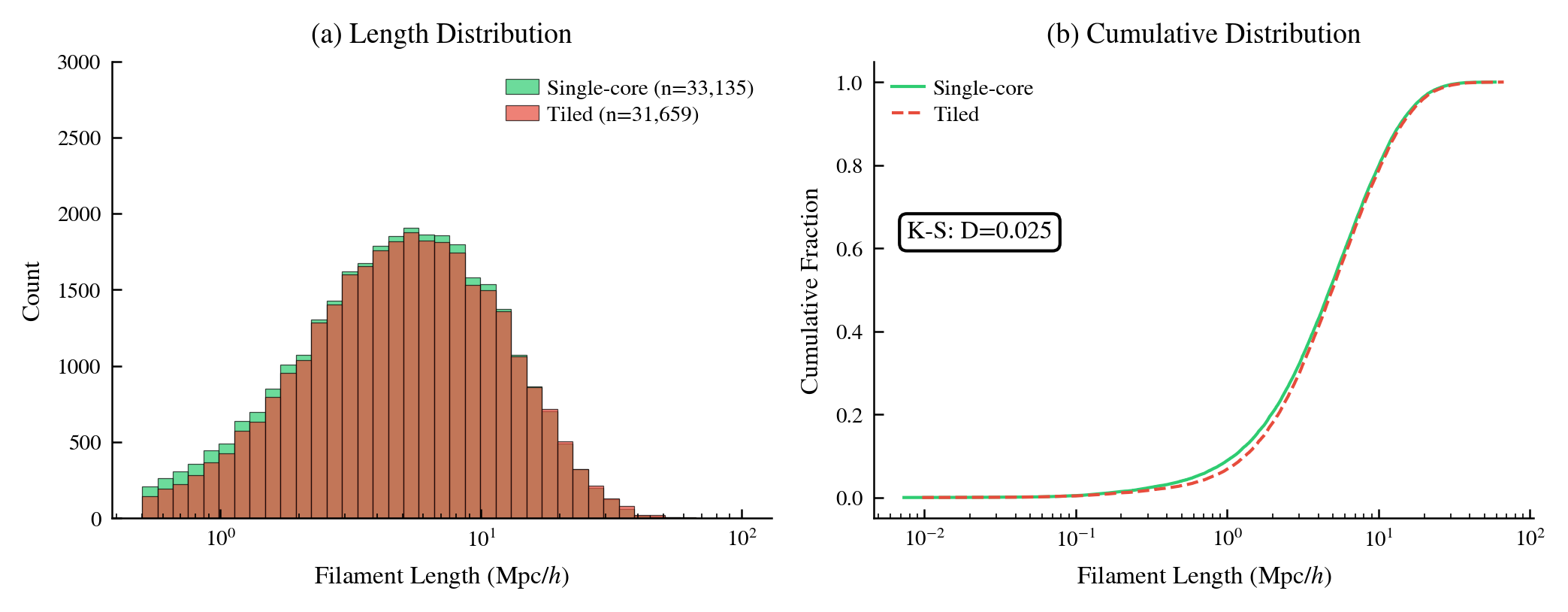}
    \caption{Filament length distribution comparison between single-core \textbf{(green)} and tiled \textbf{(red)} approaches (20-particle mass cut). \textbf{Left:} Histogram showing near-identical distributions. \textbf{Right:} Cumulative distribution confirming 99.6 per cent total length recovery. The Kolmogorov--Smirnov statistic is 0.025, indicating statistically similar distributions.}
    \label{fig:tiling_length}
\end{figure*}

\subsection{The frozen-core pipeline}
\label{sec:frozen_core}

When \disperse is run on overlapping tiles, each tile produces a complete skeleton including filaments that extend into padding regions. A naive concatenation would create:

\begin{itemize}
    \item \textbf{Duplicates:} Same filaments detected in multiple overlapping tiles
    \item \textbf{Fragments:} Filaments artificially cut at tile boundaries
    \item \textbf{Artifacts:} Spurious short filaments at edges due to incomplete topology
\end{itemize}

We implement a \emph{frozen core} processing pipeline with three stages to merge tile outputs while preserving filament integrity. As we demonstrate in Section~\ref{sec:validation}, the pipeline can be simplified to a single empirical parameter.

\subsubsection{Stage 1: Frozen Core Filtering}

Each tile has a core region (e.g., 100\,\mpch) plus padding (e.g., 60\,\mpch). For filament segments in each tile:

\begin{itemize}
    \item \textbf{Keep} if at least one endpoint is inside the core region
    \item \textbf{Discard} if both endpoints are in the padding zone
\end{itemize}

This ensures each filament is ``owned'' by the tile whose core contains its endpoint, preventing double counting while preserving filaments that straddle boundaries.

\subsubsection{Stage 2: Spatial Deduplication}

Despite frozen-core filtering, some filaments may appear in multiple tiles when both endpoints lie near tile boundaries. These duplicates arise when a filament lies entirely within the overlap region of two adjacent tiles: both tiles detect the filament independently, producing two copies with nearly identical centre positions, lengths, and sampling-point counts. Unlike fragments (which are portions of a single filament split at a tile boundary), these are complete, redundant detections of the same physical structure. We remove them using KDTree-based deduplication:

\begin{enumerate}
    \item Compute center position of each filament: $\mathbf{c}_i = \frac{1}{N}\sum_{j} \mathbf{p}_{ij}$
    \item Build spatial index (KDTree) of centers
    \item Find pairs within tolerance $\delta$
    \item For each pair, keep the filament with more sample points.
\end{enumerate}

This stage is essential: without it, the tiled catalogue contains ${\sim}$8 per cent excess total length from duplicate filaments (Table~\ref{tab:pipeline_simplification}). The algorithm has $O(n \log n)$ scaling.

\subsubsection{Stage 3: Boundary Stitching}

Filaments fragmented at tile boundaries are reconnected using the same tolerance $\delta$:
\begin{enumerate}
    \item Identify endpoints within $\delta$ of tile boundaries
    \item Match end$\rightarrow$start point pairs using KDTree, requiring that the two endpoints lie on \emph{opposite sides} of the same tile boundary to prevent spurious along-boundary connections
    \item Apply Union Find to merge connected components
\end{enumerate}

The implementation also includes two optional safeguards against artificial loop formation: a pre-filter that rejects connections where the resulting filament would close on itself (start$_i \approx$ end$_j$ within 10\,\mpch), and a post-filter that breaks stitched groups with a length-to-extent ratio exceeding 2.5. As we show in Appendix~\ref{sec:antiloop}, neither safeguard has a material effect on the results for our dataset: length recovery, maximum filament length, and the number of super-filaments are identical with or without them (Table~\ref{tab:antiloop_necessity}). The anti-loop protections are therefore not included in the default pipeline; they are available as options in the code for application to other catalogues where loop artefacts may be more prevalent.

This stage does not change the total filament length (which remains at 99.56 per cent of the reference) but reduces the filament count by reconnecting fragments that were split at tile boundaries.

Both stages use a single matching tolerance $\delta = 0.5$\,\mpch, the only empirical parameter in the post processing pipeline. As shown in Section~\ref{sec:validation}, the results are robust to the choice of $\delta$ within the range $0.1-0.7$ \mpch.

\subsubsection{Method Validation}

To demonstrate that the frozen core pipeline preserves filament structure, we compare tiled processing against a monolithic reference on a 300\,\mpch\ subvolume extracted from the centre of the MDPL2 box (avoiding periodic boundary effects). The single-core reference processes this 300\,\mpch\ MDPL2 subvolume with 60\,\mpch\ padding on each side monolithically, while the tiled approach uses $3 \times 3 \times 3 = 27$ tiles (each 100\,\mpch\ core + 60\,\mpch\ padding) merged via the frozen core pipeline. Both use identical \disperse parameters: persistence threshold $n_\sigma = 7.0$ (the number of noise sigma that a topological feature's density contrast must exceed to be retained; Section~\ref{sec:disperse_bottleneck}) and internal skeleton smoothing $s = 8$ (\texttt{skelconv -smooth 8}), which applies 8 iterations of Laplacian averaging to the raw \disperse skeleton vertices before output, reducing Delaunay-grid wiggles in the individual tile skeletons.

\begin{figure*}
    \centering
    \includegraphics[width=\textwidth]{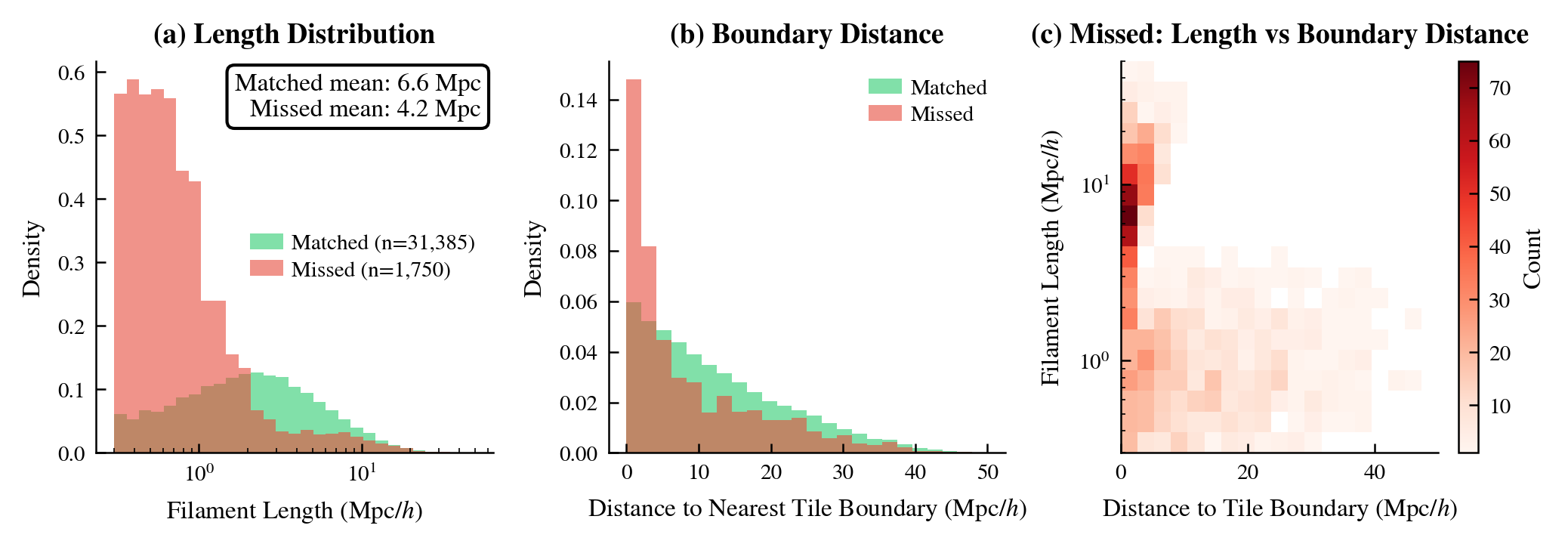}
    \caption{\textbf{Unmatched filament analysis} between single-core reference and tiled output (20-particle mass cut). \textbf{Left:} Length distribution of matched (green) vs.\ unmatched (red) filaments; unmatched filaments are systematically shorter (mean 4.2\,\mpch\ vs.\ 6.6\,\mpch). \textbf{Centre:} Distance to nearest tile boundary; long unmatched filaments cluster near boundaries, while short ones are distributed throughout. \textbf{Right:} Joint distribution of unmatched filament length and boundary distance.}
    \label{fig:tiling_unmatched}
\end{figure*}

Figure~\ref{fig:tiling_visual} shows the visual comparison: the tiled approach (centre) reproduces the filament network from the single-core reference (left) with high fidelity. The cosmic web structure, with filaments tracing high-density ridges, is preserved across the entire 300\,\mpch\ region, including near the tile boundaries (dashed white lines). The right panel isolates the discrepancies directly: of the filaments in the slice, only 60 (red) are present in the single-core catalogue but absent from the tiled output, while 11 (blue) are unique to the tiled catalogue. In this slice the discrepant filaments lie predominantly near the tile boundaries (dashed grid), confirming that the residual differences do not corrupt the interior of each tile; the full-volume analysis below (Fig.~\ref{fig:tiling_unmatched}) shows that this near-boundary concentration is driven by the longer missing filaments, while the more numerous short unmatched filaments are distributed throughout the volume.

\begin{table}
    \centering
    \caption{Frozen core tiling validation: comparison of single-core and tiled filament detection on a 300\,\mpch\ MDPL2 subvolume (20-particle mass cut).}
    \label{tab:tiling_validation}
    \small
    \begin{tabular}{lccc}
        \hline
        \textbf{Metric} & \textbf{Reference} & \textbf{Tiled} & \textbf{Recovery} \\
        \hline
        Filament count & 33,135 & 31,659 & 95.5\% \\
        Total length (\mpch) & 213,577 & 212,636 & 99.6\% \\
        Mean length (\mpch) & 6.45 & 6.72 & -- \\
        Median length (\mpch) & 4.75 & 4.95 & -- \\
        \hline
        Matched pairs & \multicolumn{2}{c}{31,385} & 94.7\% \\
        Length ratio & \multicolumn{2}{c}{1.000 (median)} & -- \\
        Centroid offset & \multicolumn{2}{c}{$0.00$\,\mpch\ (median)} & -- \\
        \hline
    \end{tabular}
\end{table}

Table~\ref{tab:tiling_validation} quantifies the recovery: the tiled approach achieves 99.6\% total length recovery and 94.7\% individual filament matching. Figure~\ref{fig:tiling_length} shows the length distributions are near identical (Kolmogorov Smirnov statistic 0.025), confirming that the frozen core pipeline preserves filament statistics.

The ${\sim}$5\% count difference arises from systematically shorter unmatched filaments: unmatched filaments have mean length 4.2\,\mpch\ vs.\ 6.6\,\mpch\ for matched filaments (Figure~\ref{fig:tiling_unmatched}). However, short unmatched filaments are not preferentially near tile boundaries, they are distributed throughout the volume, arising from tetrahedra whose circumspheres extend beyond the frozen core boundary at any location. Only long unmatched filaments cluster near tile boundaries, where missing surrounding context disrupts extended structures spanning multiple tiles. The right panel of Figure~\ref{fig:tiling_unmatched} confirms this dichotomy in the joint length--boundary-distance distribution: the long-unmatched filaments are concentrated in the corner nearest the tile boundary, while short ones populate the full range of distances. This is a conservative bias: we may miss some short filaments anywhere in the volume, but we do not introduce spurious structures.

The frozen core method also preserves the critical points of the Morse--Smale complex with high fidelity. Comparing the tiled and monolithic critical points within the 300\,\mpch\ core (20{,}136 critical points in the reference), we find 100 per cent recovery of density maxima (3{,}841 peaks) and minima (254 void centres), 99.7 per cent recovery of 2-saddles (9{,}814 of 9{,}839), and 89.4 per cent recovery of 1-saddles (5{,}544 of 6{,}202), with zero spurious critical points introduced by the tiling. The overall critical point recovery is 96.6 per cent (at a matching tolerance of 0.1\,\mpch). The perfect recovery of peaks and minima is particularly relevant, as many studies use \disperse density maxima as the basis for further analysis \citep[e.g.][]{Cohn2022DisperseClusters, GalarragaEspinosa2024, Cornwell2024}.

\subsubsection{Pipeline simplification analysis.}
\label{sec:validation}
To determine which post processing stages are necessary, we systematically test the pipeline with individual stages removed (Table~\ref{tab:pipeline_simplification}).

\emph{Deduplication is essential.} Without it, the tiled catalogue retains ${\sim}$8 per cent excess total length from duplicate filaments that appear in overlapping tiles. Of the 3{,}072 duplicate pairs identified at $\delta = 0.5$\,\mpch, 1{,}719 are cross tile pairs with identical centres, lengths, and point counts. These are genuine duplicates arising from the frozen core overlap.

\emph{Stitching improves topology but does not affect total length.} The total length recovery is 99.56 per cent with or without stitching. Stitching reconnects ${\sim}$875 filament fragments at tile boundaries, reducing the filament count and producing physically continuous structures. Without it, boundary filaments remain fragmented but no information is lost.

The simplified pipeline therefore has a single empirical parameter $\delta = 0.5$\,\mpch, used for both deduplication (center matching) and stitching (endpoint matching). We validate the robustness of this choice in Appendix~\ref{sec:parameter_sensitivity}, varying $\delta$ over $0.05-5.0$ \mpch. Results are stable for $\delta \leq 0.7$\,\mpch: length recovery exceeds 99.4 per cent and stitching does not create artificial super filaments. We also tested anti-loop protections during stitching (pre filter on loop distance and post filter on length/extent ratio) and found them unnecessary (Appendix~\ref{sec:antiloop}).

\begin{table*}
\centering
\caption{Pipeline simplification analysis. Each row shows the effect of removing a post processing stage. Deduplication is essential (removes ${\sim}$8\% excess length); stitching improves topology without affecting total length. The simplified pipeline uses a single parameter $\delta = 0.5$\,\mpch. $N_\mathrm{super}$ counts ``super filaments'': tiled filaments whose length exceeds the maximum filament length in the single-core reference run (58.8\,\mpch) by more than 0.01\,\mpch, indicating spurious over-stitching or incomplete deduplication. The 0.01\,\mpch\ threshold excludes floating-point level differences between the tiled and single-core tessellations.}
\label{tab:pipeline_simplification}
% \resizebox{\columnwidth}{!}{%
\begin{tabular}{lccccc}
\hline\hline
Configuration & Params & $N_\mathrm{fil}$ & Length rec.\ (\%) & Max $L$ (\mpch) & $N_\mathrm{super}$ \\
\hline
Filter only (no post processing) & 0 & 34,836 & 108.44 & 58.8 & 0 \\
Dedup only & 1 & 32,030 & 99.56 & 58.8 & 0 \\
Dedup + stitch & 1 & 31,155 & 99.56 & 67.1 & 6 \\
Stitch only (no dedup) & 1 & 33,456 & 108.44 & 113.2 & 26 \\
\hline\hline
\end{tabular}%
\end{table*}

Together, these tests confirm that the frozen core method preserves the \disperse filament network with $<$1 per cent systematic uncertainty in total filament length, at the cost of a conservative ${\sim}$5 per cent loss of short filaments near tile boundaries.

\section{Results}
\label{sec:results}

We apply the frozen-core pipeline to the full 1\,\gpc \ MDPL2 box (92 million haloes; Section~\ref{sec:mdpl2}), tiled into $10 \times 10 \times 10 = 1000$ blocks of 100\,\mpch\ core with 60\,\mpch\ padding. The Delaunay tessellation of each padded block (${\sim}220$\,\mpch, containing ${\sim}1$ million haloes) requires ${\sim}5$--$10$\,GB of memory, well within the capacity of standard compute nodes. Below we characterise the resulting filament catalogue.

\begin{figure*}
\centering
\includegraphics[width=\textwidth]{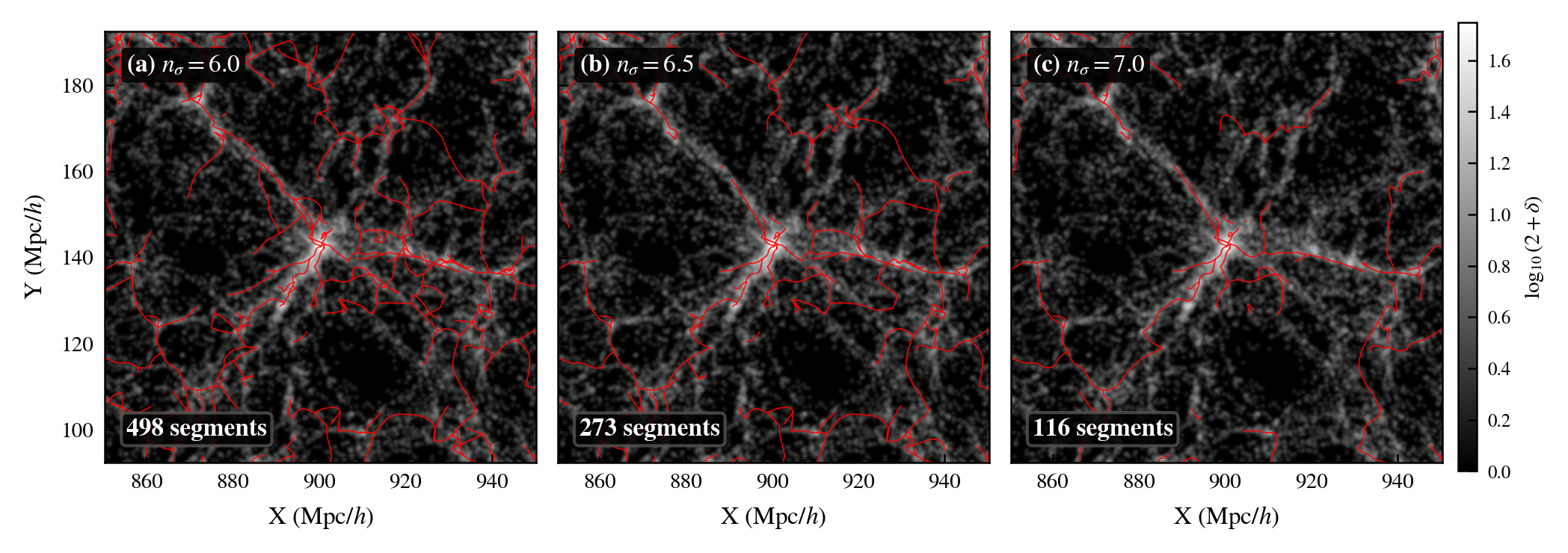}
\caption{Overlay of the filament network at three persistence thresholds for cluster number 6 from The Three Hundred project \citep[][also shown in Fig.~\ref{fig:tiling_visual}]{Cui2018}. The background shows the galaxy overdensity field $\log_{10}(2+\delta)$ in greyscale (light = overdense). Higher thresholds retain only the most prominent filaments; the large-scale network topology is preserved across all three.}
\label{fig:nsigma_slices} 
\end{figure*}

\subsection{Sensitivity to Persistence Threshold}
\label{sec:nsigma}

The persistence threshold $n_\sigma$ controls which topological features are retained in the filament catalogue. Lower thresholds preserve more structures, including low significance features that may arise from noise in the density field, while higher thresholds retain only the most persistent features. We adopt $n_\sigma = 6.5$ as our fiducial threshold, validated by matching \disperse density maxima against the MDPL2 halo catalogue (Appendix~\ref{sec:cp_halo_calibration}). To characterize the dependence of the filament network on this parameter, we run the full frozen core pipeline on the 1\,\gpc \ MDPL2 box at three persistence thresholds: $n_\sigma = 6.0$, $6.5$, and $7.0$, using identical tiling, padding, and post processing parameters throughout.

Table~\ref{tab:nsigma_comparison} summarises the filament statistics for each threshold. Increasing $n_\sigma$ from 6.0 to 7.0 reduces the filament count by a factor of 4.7, from $2.29 \times 10^6$ to $4.86 \times 10^5$. The total filament length decreases by a factor of 2.3 over the same interval, from $1.11 \times 10^7$\,\mpch\ to $4.74 \times 10^6$\,\mpch. The less steep decline in total length compared to filament count reflects the preferential removal of short filaments: the mean filament length increases from 4.85\,\mpch\ at $n_\sigma = 6.0$ to 9.77\,\mpch\ at $n_\sigma = 7.0$, while the median increases from 3.65 to 7.37\,\mpch. The width of the distribution also broadens, with the standard deviation growing from 4.16 to 8.60\,\mpch.

\begin{table}
\centering
\caption{Filament catalogue statistics for three persistence thresholds applied to the 1\,\gpc\ MDPL2 box (92 million haloes after 20-particle mass cut, frozen core tiling with 60\,\mpch\ padding).}
\label{tab:nsigma_comparison}
\begin{tabular}{lccc}
\hline\hline
Statistic & $n_\sigma = 6.0$ & $n_\sigma = 6.5$ & $n_\sigma = 7.0$ \\
\hline
Filaments & 2,288,633 & 1,155,755 & 485,703 \\
Total length (\mpch) & 11,098,226 & 7,532,813 & 4,744,234 \\
Mean length (\mpch) & 4.85 & 6.52 & 9.77 \\
Median length (\mpch) & 3.65 & 4.88 & 7.37 \\
Std dev (\mpch) & 4.16 & 5.71 & 8.60 \\
Max length (\mpch) & 75.58 & 82.30 & 130.66 \\
\hline\hline
\end{tabular}
\end{table}

Figure~\ref{fig:nsigma_slices} shows an overlay of the filament network at the cluster number 6 from the Three Hundred simulation (also shown in Fig.~\ref{fig:tiling_visual}). The background shows the galaxy overdensity field $\log_{10}(2+\delta)$ in greyscale (light = overdense). At $n_\sigma = 6.0$, the network is densely populated, with filaments tracing both prominent inter-cluster bridges and tenuous structures in lower-density regions. At $n_\sigma = 6.5$, the primary filamentary backbone of the cosmic web remains intact while finer structures are removed. At $n_\sigma = 7.0$, only the most prominent filaments persist, corresponding to the ridges connecting the most massive nodes of the cosmic web. The large-scale topology is preserved across all three thresholds, while the filling factor of the network changes substantially.

\begin{figure*}
\centering
\includegraphics[width=\textwidth]{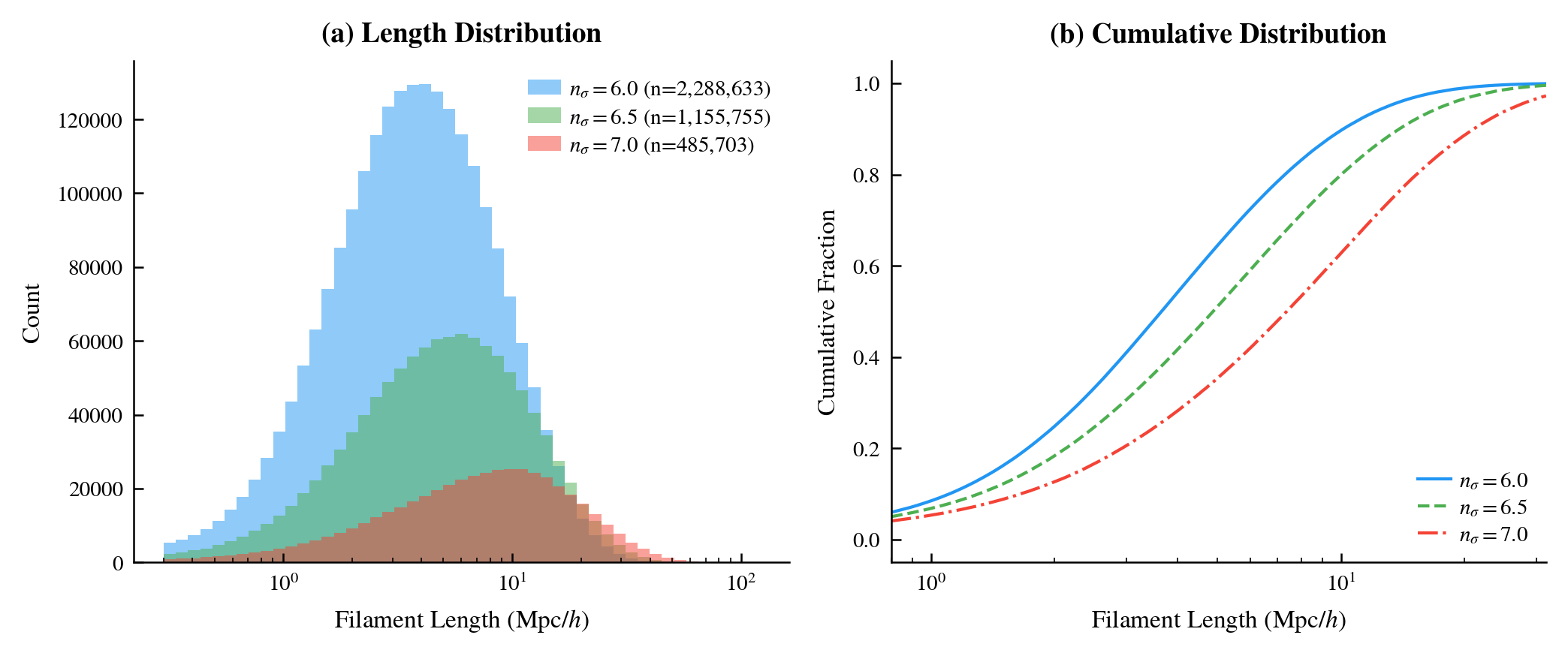}
\caption{Filament length distributions for the three persistence thresholds. (a)~Histogram showing the count distribution on a logarithmic length axis; the total count decreases and the peak shifts to longer filaments with increasing $n_\sigma$. (b)~Cumulative fraction showing the broadening of the distribution toward longer filaments at higher thresholds; the x-axis is limited to ${\sim}32$\,\mpch\ where the cumulative fraction plateaus.}
\label{fig:nsigma_lengths}
\end{figure*}

The length distribution (Figure~\ref{fig:nsigma_lengths}a) shifts systematically with $n_\sigma$. At $n_\sigma = 6.0$ the histogram peaks at $L \approx 4.2$\,\mpch, rising to $6.0$\,\mpch\ at $n_\sigma = 6.5$ and $9.8$\,\mpch\ at $n_\sigma = 7.0$. The distribution declines steeply at larger lengths. The cumulative distribution (Figure~\ref{fig:nsigma_lengths}b) illustrates the broadening: 80 per cent of filaments are shorter than $7.3$\,\mpch\ at $n_\sigma = 6.0$, whereas the same fraction is reached at approximately $15.3$\,\mpch\ for $n_\sigma = 7.0$.

\begin{figure*}
\centering
\includegraphics[width=\textwidth]{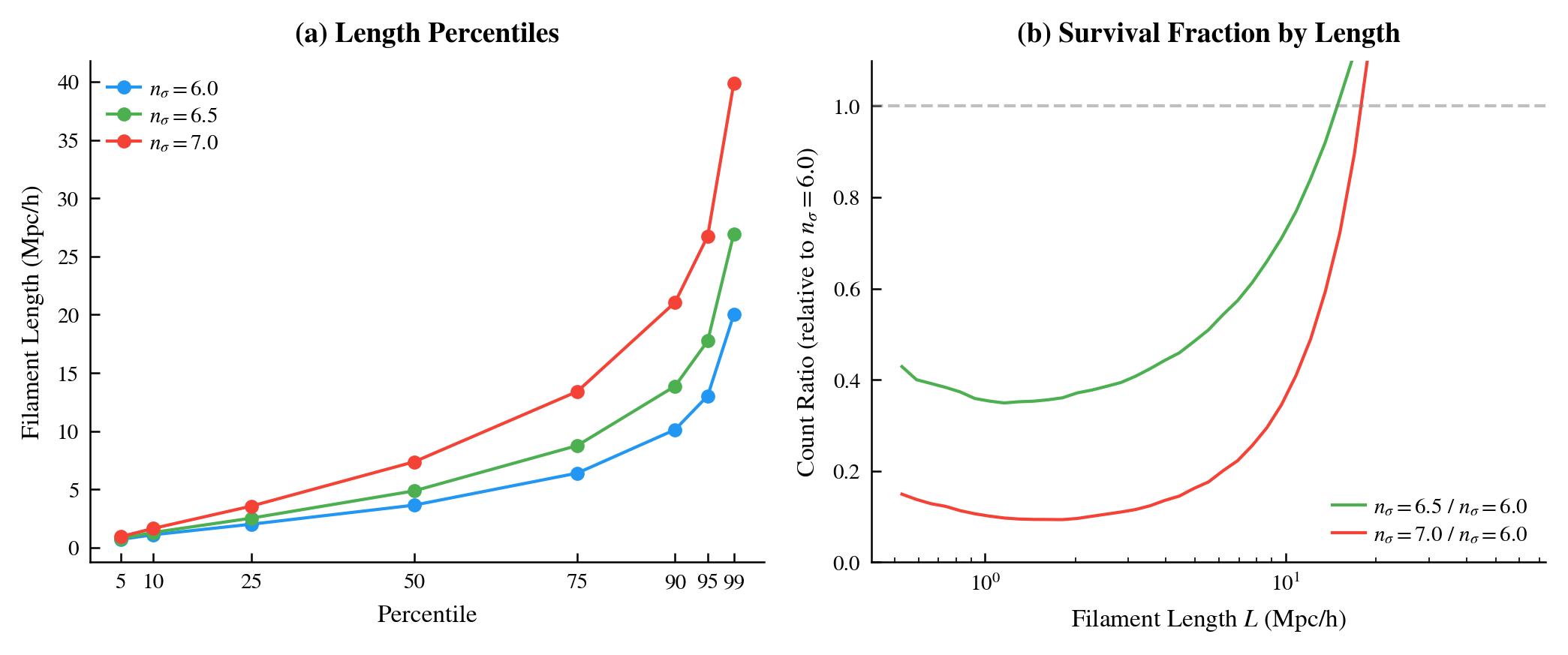}
\caption{(a)~Length percentiles for each persistence threshold; the distribution shifts to longer lengths at higher $n_\sigma$ across all percentiles. (b)~Count ratio relative to $n_\sigma = 6.0$ as a function of filament length. Short filaments ($L < 2$\,\mpch) retain only ${\sim}$10 per cent of their population at $n_\sigma = 7.0$, while the ratio exceeds unity for $L \gtrsim 20$\,\mpch\ as persistence pair cancellation merges shorter arcs into longer structures.}
\label{fig:nsigma_survival}
\end{figure*}

Figure~\ref{fig:nsigma_survival}a shows the filament length percentiles at each threshold: every percentile, from the 10th to the 90th, shifts to larger lengths as $n_\sigma$ increases, confirming that the distribution broadens uniformly rather than being driven by the tails alone. To examine which filament lengths are most affected by the threshold, Figure~\ref{fig:nsigma_survival}b shows the ratio of filament counts at $n_\sigma = 6.5$ and $7.0$ to those at $n_\sigma = 6.0$ as a function of length. At $n_\sigma = 7.0$, the count ratio is approximately 0.10 for filaments shorter than 2\,\mpch, rising to above unity for filaments longer than ${\sim}20$\,\mpch. This length dependent sensitivity is consistent with the expectation from persistence theory: longer filaments tend to connect higher contrast density features, producing larger persistence values that exceed more stringent thresholds, and the cancellation of persistence pairs at intermediate saddle points merges shorter arcs into longer continuous structures. The maximum filament length increases from 75.6\,\mpch\ at $n_\sigma = 6.0$ to 130.7\,\mpch\ at $n_\sigma = 7.0$. In \disperse, increasing $n_\sigma$ cancels persistence pairs below the threshold, removing saddle points and merging the filament segments that connected at those critical points into single longer structures \citep{2011aSousbie}. This mechanism accounts for the emergence of longer filaments at higher thresholds.

The filament number density decreases from $2.29 \times 10^{-3}$\,(\mpch)$^{-3}$ at $n_\sigma = 6.0$ to $4.86 \times 10^{-4}$\,(\mpch)$^{-3}$ at $n_\sigma = 7.0$, while the total filament length per unit volume decreases from $1.11 \times 10^{-2}$ to $4.74 \times 10^{-3}$\,(\mpch)$^{-2}$ over the same range. These densities provide a quantitative characterisation of the cosmic web network at different significance levels in the largest volume analysed to date with topologically rigorous methods, and can serve as a reference for comparison with other filament finders and cosmological simulations.

\begin{figure*}
\centering
\includegraphics[width=0.85\textwidth]{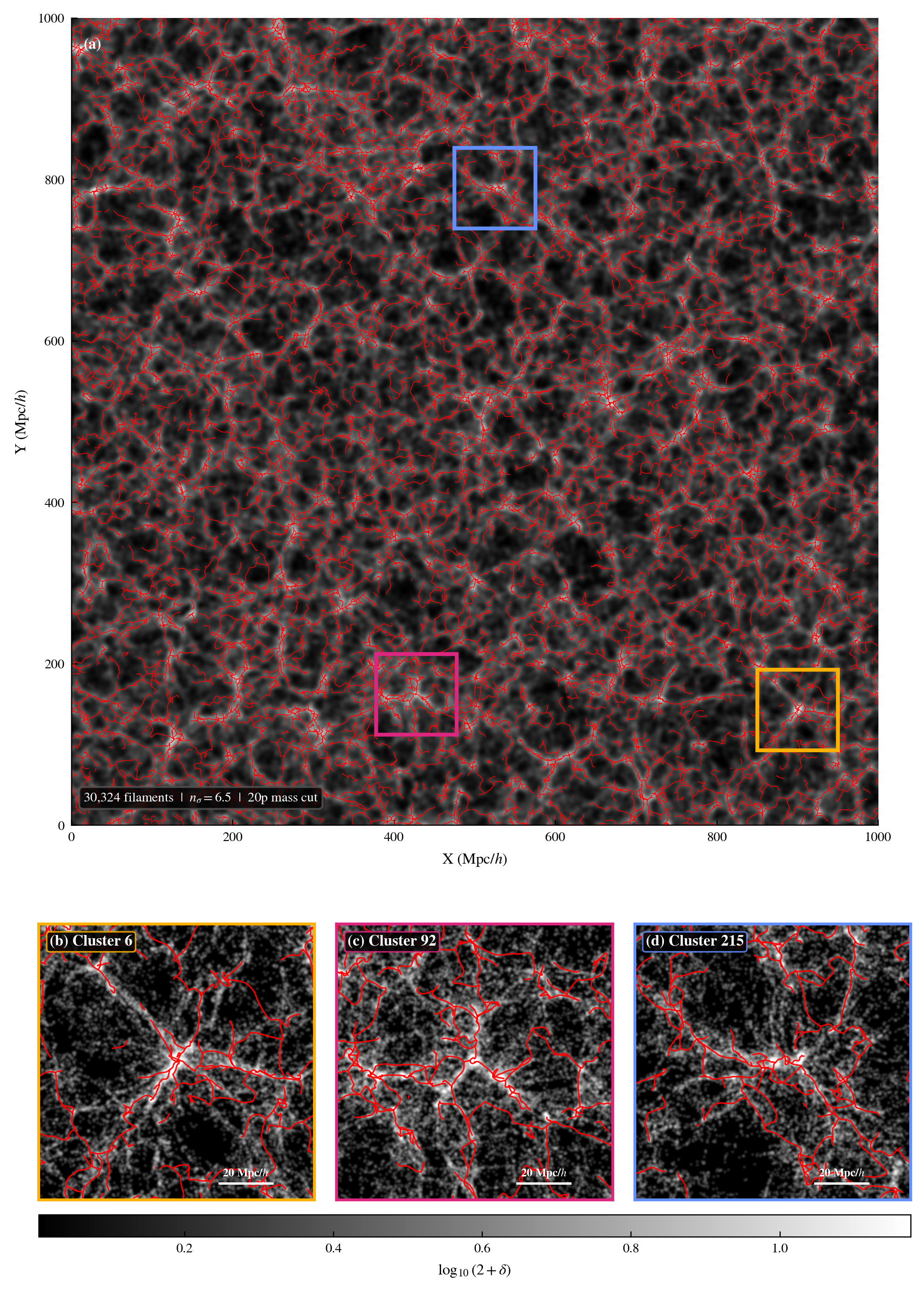}
\caption{The filament network in the 1\,\gpc\ MDPL2 simulation at $n_\sigma = 6.5$. \textit{Top:} Galaxy overdensity field (colour scale) in a 20\,\mpch\ thick projection at $z = 950$\,\mpch, with filaments overlaid. Coloured boxes mark regions shown in the inset panels. \textit{Bottom:} Zoom into 100\,\mpch\ regions around three clusters from The Three Hundred project \citep{Cui2018}, showing the convergence of filaments toward the most massive nodes of the cosmic web. The colourbar shows $\log_{10}(2 + \delta)$, where $\delta$ is the galaxy overdensity. The filament network is periodic, inheriting the periodic boundary conditions of the MDPL2 box.}
\label{fig:cluster_filaments}
\end{figure*}

\subsection{The Gigaparsec-Scale Filament Network}
\label{sec:filament_network}

Figure~\ref{fig:cluster_filaments} presents a slice of the full 1\,\gpch\ filament network detected in the MDPL2 simulation at $n_\sigma = 6.5$. The main panel shows the projected galaxy overdensity field $\delta = \rho/\bar{\rho}  - 1$ (displayed as $\log_{10}(2 + \delta)$) in a 20\,\mpch thick slice at $Z = 950$\,\mpch, with ${\sim}30{,}000$ filaments overlaid. The cosmic web structure is clearly visible, with filaments tracing the overdense ridges connecting massive haloes and delineating the boundaries of voids. The filament network is periodic, inheriting the periodic boundary conditions of the MDPL2 box: filaments that reach a face of the volume continue across the opposite face, and the minimum-image convention is applied throughout the subsequent post-processing and connectivity analysis.

The three inset panels zoom into 100\,\mpch\ regions centred on clusters from The Three Hundred project \citep{Cui2018}, which correspond to the 324 most massive haloes in MDPL2 ($M_\mathrm{vir} > 8 \times 10^{14}\,\msun\,h^{-1}$). Each inset illustrates the filamentary connectivity of a massive cluster node: multiple filaments converge radially toward the cluster centre, forming the dense network of inter cluster bridges characteristic of the cosmic web. These structures are recovered robustly by the frozen-core tiling pipeline despite being distributed across multiple tile boundaries.

The filaments identified in this work are large-scale skeletal structures traced by the dark matter halo distribution. At our fiducial threshold $n_\sigma = 6.5$, the typical filament has a length of ${\sim}5$--$7$\,\mpch\ (median 4.9\,\mpch, mean 6.5\,\mpch; Table~\ref{tab:nsigma_comparison}), set by the inter-node spacing of the cosmic web at this tracer density. The mean inter-halo separation of ${\sim}2.2$\,\mpch\ determines the effective spatial resolution: structures on scales smaller than this are not resolved. Because these arcs span the inter-node distance with little curvature (median post-smoothing sinuosity $1.011$; Section~\ref{sec:connectivity_postprocessing}), they are well approximated as near-straight bridges between neighbouring density maxima. These large-scale filaments are expected to host the WHIM bridges between clusters and groups identified in hydrodynamical simulations \citep{CenOstriker1999, Dave2001}, which have characteristic overdensities of $\delta \sim 1$--$100$ and temperatures of $10^5$--$10^7$\,K, although we do not measure the gas properties directly here. The frozen-core method would apply equally to higher-density tracer populations (e.g.\ galaxy catalogues with ${\sim}10\times$ higher number density), where the smaller mean inter-tracer separation would produce shorter circumradii and hence require smaller padding widths; conversely, very sparse tracers would need larger padding. The circumradius distribution should be re-evaluated for each new application (Section~\ref{sec:padding}).

\subsection{Cluster Connectivity}
\label{sec:connectivity}

The connectivity $\kappa$ of a node, i.e., the number of filaments attached to a cluster or group, has emerged as a key topological diagnostic of the cosmic web. Early $N$-body work by \citet{Colberg2005Intercluster} showed that more massive clusters are connected to more filaments, with the most massive clusters averaging nearly five filamentary connections. \citet{Codis2018} derived analytically from Gaussian random field theory that three-dimensional nodes have a mean connectivity of $\kappa \approx 6.11$ (exact from the saddle-to-maximum ratio in 3D). Using cosmological simulations, the same work showed that connectivity grows logarithmically with halo mass, well fitted by $\kappa(M) \approx \frac{10}{3}\log_{10}(M/10^{11}\,\msun)$. This positive mass-connectivity relation has been confirmed observationally in the COSMOS survey \citep{DarraghFord2019} and in photometric cluster samples \citep{Sarron2019}, and in hydrodynamical simulations using both the IllustrisTNG simulation \citep{Gouin2021} and The Three Hundred project \citep{Santoni2024}. Connectivity has been shown to correlate with cluster dynamical state and formation history \citep{Gouin2021}, making it a probe of mass assembly beyond what is captured by halo mass alone. Here we present the first measurement of cluster connectivity from halo-traced filaments across the full 1\,\gpch MDPL2 volume.

\subsubsection{Cluster sample}

We measure connectivity for the 324 most massive haloes in MDPL2, which are the parent haloes of The Three Hundred project \citep{Cui2018}. These clusters span a virial mass range $M_\mathrm{vir} = 8.0 \times 10^{14}$--$3.5 \times 10^{15}\,h^{-1}\,\msun$. Positions and masses are taken directly from the Rockstar halo catalogue at $z = 0$. For each cluster we extract $M_{200\mathrm{c}}$ from the Rockstar catalogue and compute
\begin{equation}
R_{200} = \left( \frac{3\,M_{200\mathrm{c}}}{4\pi \times 200\,\rho_\mathrm{crit}} \right)^{1/3},
\label{eq:r200}
\end{equation}
where $\rho_\mathrm{crit} = 2.775 \times 10^{11}\,h^{-1}\,\msun\,(h^{-1}\,\mathrm{Mpc})^{-3}$ is the critical density at $z = 0$ in simulation units ($h$-scaled mass and length). This yields $R_{200}$ values in the range $1.64$--$2.94$\,\mpch. To test whether the mass-connectivity scaling extends to lower masses, we additionally draw an extended sample of ${\sim}22{,}900$ haloes spanning $M_{200\mathrm{c}} = 10^{12}$--$10^{15.5}\,h^{-1}\,\msun$ (2{,}000 haloes drawn randomly per 0.25-dex mass bin) from the full 1\,\gpc\ filament catalogue. Since the sample is binned by mass before fitting, the random selection within each bin does not bias the mass--connectivity relation; it only affects the statistical uncertainty within each bin, which is captured by the bootstrap error bars. This broader sample is analysed in Section~\ref{sec:connectivity_results}.

\subsubsection{Filament post-processing}
\label{sec:connectivity_postprocessing}

Before measuring connectivity, we apply a five-step topological cleaning pipeline to the raw \disperse\ output. This pipeline follows the standard practices established in the literature \citep{Malavasi2020, GalarragaEspinosa2020} and is applied independently at each persistence threshold. The same pipeline is used in a companion study (Singh et al., in preparation) that classifies galaxy groups by their proximity to cluster-connected filaments to investigate channelled versus isotropic accretion onto clusters, ensuring consistency between the two studies.

\begin{enumerate}
    \item \textbf{Zero-length filter} (threshold $< 0.01$\,\mpch): removes near-zero-length arcs that arise as topological defects from Poisson noise in the halo density field. These saddle-maximum pairs persist even after persistence simplification and have no physical counterpart \citep{Malavasi2020}.

    \item \textbf{Isolated maxima filter}: removes density maxima connected to only one filament, together with their single connected arc. Degree-1 nodes are non-physical in the cosmic web topology, where every node should be connected to at least two filaments \citep{Malavasi2020}.

    \item \textbf{Length filter} (threshold $< 1.0$\,\mpch): removes filament arcs shorter than 1\,\mpch, below the effective resolution of the halo-traced density field.

    \item \textbf{Filament merging} (angle threshold $60^\circ$, endpoint tolerance $0.5$\,\mpch): \disperse outputs individual arcs between consecutive critical points via the \texttt{-breakdown} parameter (each arc runs from a maximum to a 2-saddle or vice versa). A single physical filament connecting two clusters may therefore be split into several short arcs at intermediate saddle points. The merging step reconnects nearly collinear arcs that meet at shared saddle points, producing continuous structures that extend from one density maximum to another. The $60^\circ$ angle threshold prevents merging arcs that meet at sharp angles, which would indicate genuinely distinct filaments converging at a node rather than segments of a single structure. This step reduces the raw arc count from ${\sim}1{,}156{,}000$ to ${\sim}824{,}000$ filaments (at $n_\sigma = 6.5$; analogous reductions occur at other thresholds). The merging preserves all sampling points and the total network length; it changes only the filament count and the assignment of arcs to logical filaments. Filament endpoints after merging coincide with density maxima (the nodes of the cosmic web), so their positions are physically meaningful and are not shifted by this step.

    \item \textbf{Smoothing} (10 iterations): iterative Laplacian averaging of interior sampling points (each vertex moved to the midpoint of its two neighbours; endpoints are held fixed to preserve the critical-point positions). The raw \disperse skeleton follows the edges of Delaunay tetrahedra, producing zigzag paths that do not reflect the smooth density ridges of the cosmic web. Although \disperse was run with \texttt{skelconv -smooth 8} on each tile individually, this per-tile smoothing leaves a residual median sinuosity of $1.057$ in the stitched network. The sinuosity is defined as the ratio of the filament path length to the straight-line distance between its two endpoints; a value of 1.0 corresponds to a perfectly straight filament, so the residual sinuosity of 1.057 means filament paths are 5.7 per cent longer than the direct endpoint distance. Applying 10 further iterations post-stitching reduces the median sinuosity to $1.011$ (filament paths only 1.1 per cent longer than the endpoint distance) and decreases arc lengths by ${\sim}5.5$ per cent on average, eliminating Delaunay-grid wiggles that survive tile-level processing. The length reduction is almost entirely due to straightening artificial zigzags rather than altering the physical path of the filament. The low residual sinuosity ($1.011$) indicates that, on these large scales, the skeletal filaments are close to straight-line connections between the density maxima they link, consistent with their interpretation as the tidal bridges between rare peaks \citep{Bond1996}. Since endpoints are fixed, distances from objects to filament endpoints are unaffected by the smoothing, and the total network topology (connectivity, critical-point positions) is unchanged. The marginal improvement saturates beyond ${\sim}10$ iterations (${\lesssim}0.3$ per cent per additional iteration), making 10 a natural stopping point. The minimum-image convention is applied so that filaments crossing the periodic boundaries of the simulation volume are smoothed correctly.
\end{enumerate}

\subsubsection{Connectivity measurement}

We define $\kappa$ as the number of distinct filaments whose sampling points cross a spherical shell of radius $R$ centred on the cluster, following the methodology of \citet{Santoni2024}. Sampling points are the ordered sequence of three-dimensional coordinates that \disperse uses to represent each filament arc; they trace the gradient-flow path through the Delaunay complex and are supplemented by interpolation during the \texttt{skelconv} smoothing step. For each cluster, we pre-select all filaments in the post-processed catalogue that have at least one sampling point within $3\,R_{200}$ of the cluster centre (a generous pre-selection radius). For each such cluster--filament pair, we walk along consecutive sampling points $(p_i, p_{i+1})$ and compute the distance of each point to the cluster centre. A filament contributes to $\kappa$ if at least one pair of consecutive points straddles the shell, i.e.\ one point lies inside and the other outside $R$. Each such filament increments $\kappa$ by exactly one, regardless of how many times it crosses the shell; a curved filament that exits and re-enters the sphere is counted once. Because the merging step~(iv) combines \disperse \texttt{-breakdown} arcs into single continuous filaments before the connectivity measurement, no two filaments in the post-processed catalogue share sampling points, and the double-counting of overlapping arcs is avoided.

We measure $\kappa$ at three apertures: $1.5\,R_{200}$ \citep[following][]{DarraghFord2019}, $R_{200}$ (for direct comparison with the literature), and a fixed aperture of 1\,\mpch\ \citep[cf.][]{2010Aragon}. We report primary results at $1.5\,R_{200}$. We repeat the measurement at three persistence thresholds ($n_\sigma = 6.0$, 6.5, and 7.0) to assess the sensitivity of connectivity to the filament detection significance level.

\subsubsection{Results}
\label{sec:connectivity_results}

\begin{figure*}
\centering
\includegraphics[width=\textwidth]{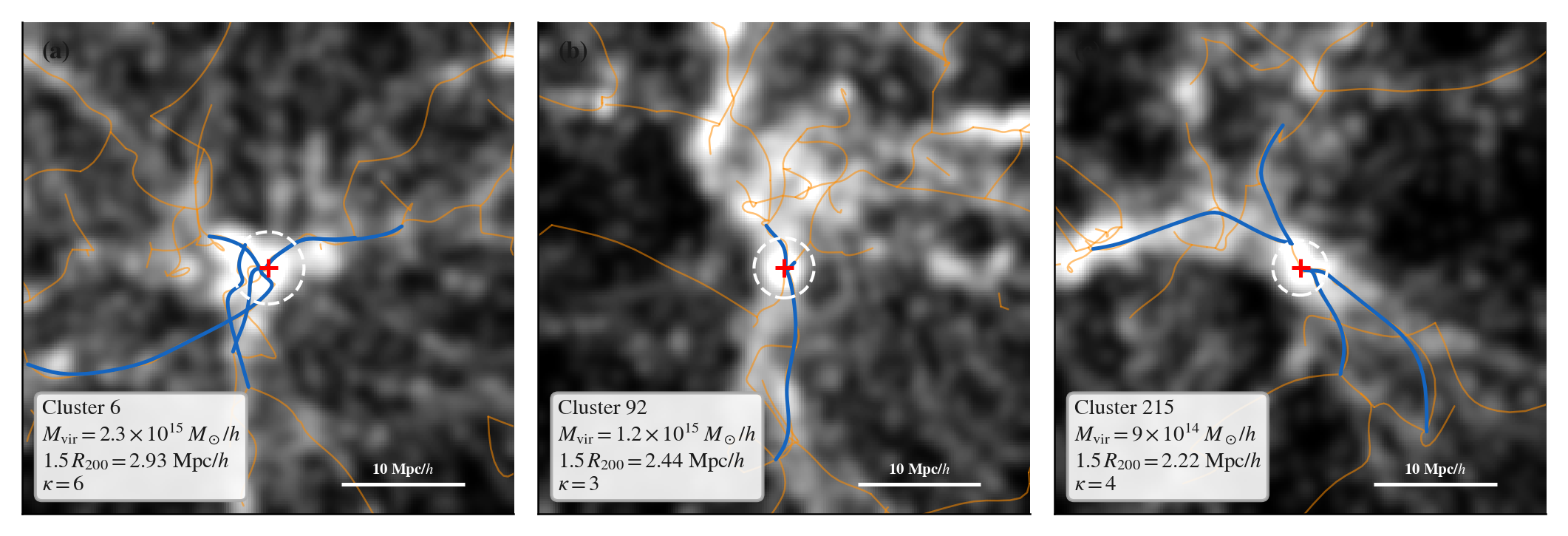}
\caption{\textbf{Connectivity illustration} for three example clusters at $n_\sigma = 6.5$, each in a $40 \times 40$\,\mpch\ region. Orange lines: background filaments within $\pm 10$\,\mpch\ of the cluster's $z$-coordinate. Blue lines: arcs crossing the $1.5\,R_{200}$ shell (dashed circle) in the post-processed catalogue. The red cross marks the cluster centre. Note that $\kappa$ is measured in three dimensions, whereas each panel shows only a $\pm 10$\,\mpch\ projected slice, so the number of connecting filaments visible by eye need not equal the annotated $\kappa$. Virial masses (as defined in \citealt{BryanNorman1998}), shell radii, and connectivities $\kappa$ are annotated. The most massive cluster (left) has the highest connectivity, consistent with the mass-connectivity relation.}
\label{fig:connectivity_illustration}
\end{figure*}

Figure~\ref{fig:connectivity_illustration} illustrates the connectivity measurement for three representative clusters spanning the mass range of our sample. Orange filaments (low opacity) show the full background network at $n_\sigma = 6.5$, while blue filaments highlight the arcs crossing the $1.5\,R_{200}$ shell; their count gives $\kappa$. The red cross marks the cluster centre. The most massive of the three clusters (Cluster 6, $M_\mathrm{vir} = 2.3 \times 10^{15}\,h^{-1}\,\msun$) is connected to $\kappa = 6$ filaments, while Cluster~92 ($M_\mathrm{vir} = 1.2 \times 10^{15}\,h^{-1}\,\msun$) has $\kappa = 3$ and Cluster~215 ($M_\mathrm{vir} = 9.2 \times 10^{14}\,h^{-1}\,\msun$) has $\kappa = 4$. While the most massive cluster has the highest connectivity, the ordering of the two lower-mass clusters is reversed relative to their masses, illustrating the substantial halo-to-halo scatter in $\kappa$ at fixed mass; the mass-connectivity relation emerges only in the statistical average over many clusters (Section~\ref{sec:connectivity_results}). We caution that the $\kappa$ values are measured in three dimensions, whereas each panel of Figure~\ref{fig:connectivity_illustration} shows only a thin projected slice, so the number of connecting filaments visible by eye need not equal the measured $\kappa$.

\begin{figure}
\centering
\includegraphics[width=\columnwidth]{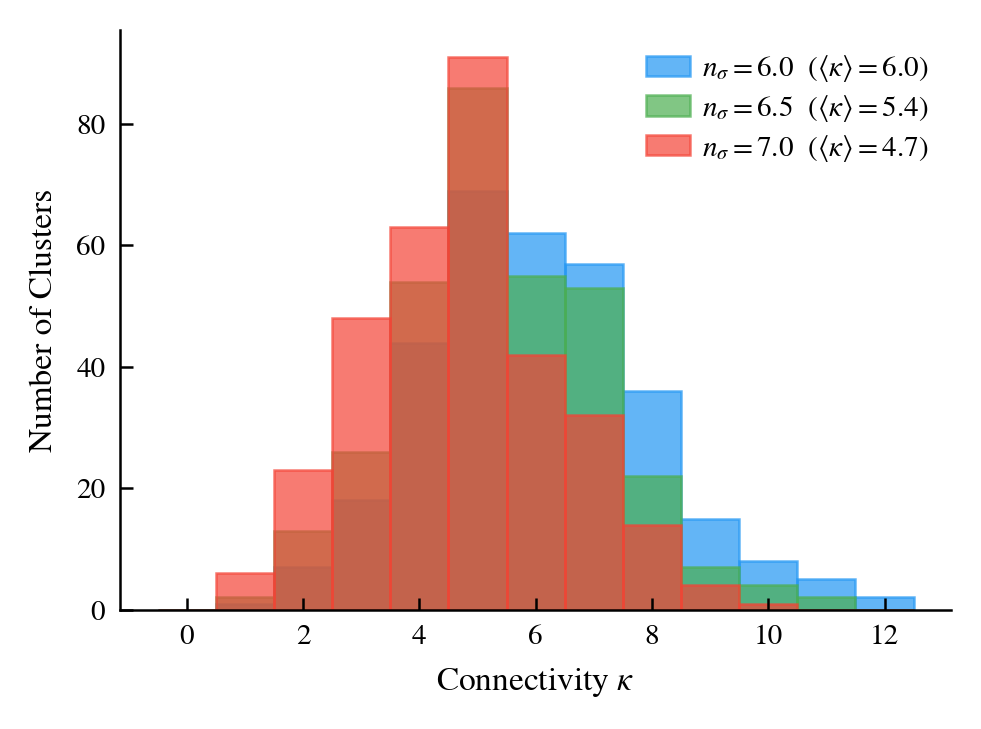}
\caption{Distribution of connectivity $\kappa$ measured at $1.5\,R_{200}$ for the 324 most massive clusters in MDPL2 at three persistence thresholds, after the five-step topological cleaning pipeline (including merging and 10-iteration smoothing). The mean connectivity decreases from $\langle \kappa \rangle = 6.0$ at $n_\sigma = 6.0$ to $\langle \kappa \rangle = 4.7$ at $n_\sigma = 7.0$.}
\label{fig:connectivity_hist}
\end{figure}

Figure~\ref{fig:connectivity_hist} shows the distribution of $\kappa$ measured at $1.5\,R_{200}$ for the three persistence thresholds. At $n_\sigma = 6.0$, the mean connectivity is $\langle \kappa \rangle = 6.0$ with a median of 6 and a range of 1--12. Increasing the threshold to $n_\sigma = 7.0$ reduces the mean to $\langle \kappa \rangle = 4.7$ (median 5, range 1--10), reflecting the removal of low-persistence filaments. The distribution shifts to lower $\kappa$ at higher thresholds while retaining the same overall shape.

\begin{figure*}
\centering
\includegraphics[width=\textwidth]{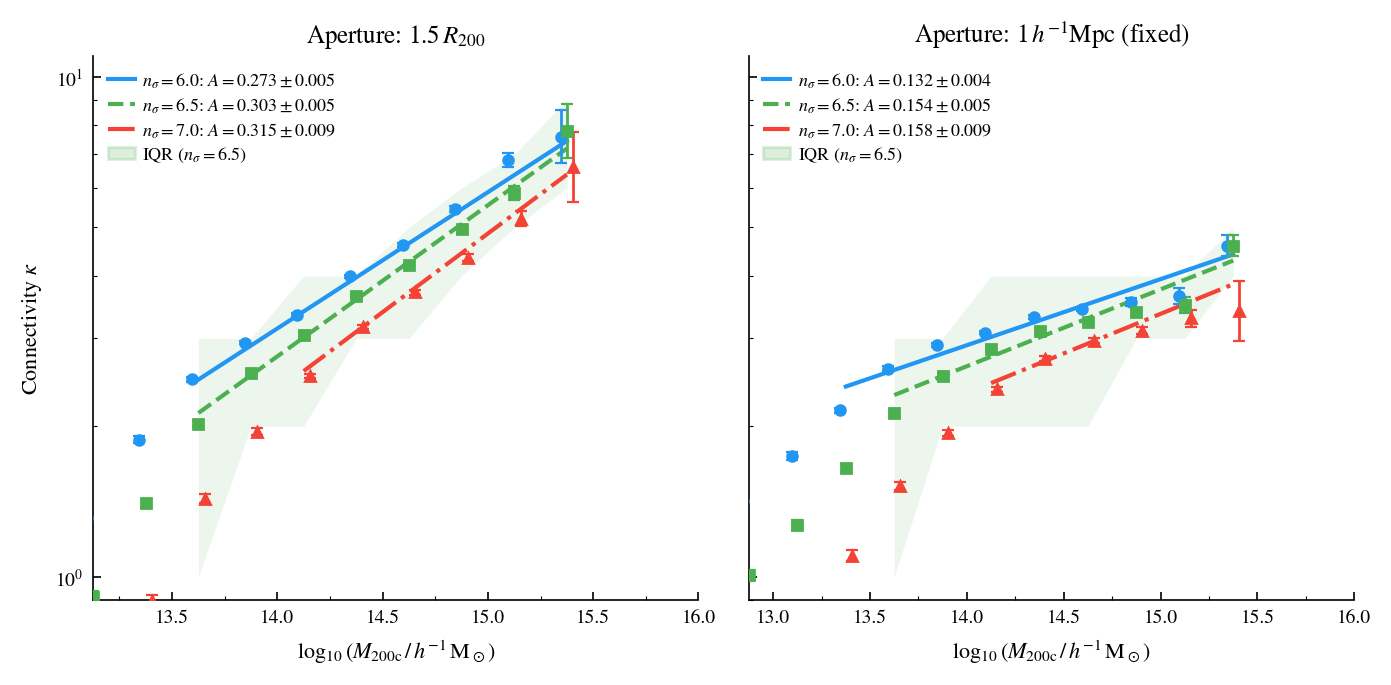}
\caption{Mass-connectivity relation for ${\sim}22{,}900$ haloes ($M_{200\mathrm{c}} = 10^{12}$--$10^{15.5}\,h^{-1}\,\msun$), measured on the 20-particle mass cut filament catalogue. \textbf{Left:} Aperture $1.5\,R_{200}$. \textbf{Right:} Fixed aperture $1\,h^{-1}\,\mathrm{Mpc}$. Large symbols show the binned means (0.25\,dex bins) for the three persistence thresholds; error bars show bootstrap $1\sigma$ confidence intervals on the mean (1{,}000 resamples per bin). The shaded band shows the interquartile (25th--75th percentile) range of the individual per-halo $\kappa$ values at the fiducial threshold $n_\sigma = 6.5$, indicating the halo-to-halo scatter in the relation; like the fits, it is drawn only over the bins with mean $\kappa > 2$, the mass range where $\kappa$ traces genuine connectivity rather than filament endpoints, and so does not extend to the lower-mass haloes that lie below the fitted relation. Lines show power-law fits to bins with mean $\kappa > 2$ (excluding unphysical filament endpoints), weighted by the inverse bootstrap variance. The slope is broadly stable across persistence thresholds at $A \approx 0.273$--$0.315$ for the $1.5\,R_{200}$ aperture.}
\label{fig:connectivity_mass}
\end{figure*}

Figure~\ref{fig:connectivity_mass} presents the mass-connectivity relation. To test whether the scaling extends beyond the cluster mass scale, we measure $\kappa$ for a broader halo sample spanning $M_{200\mathrm{c}} = 10^{12}$--$10^{15.5}\,h^{-1}\,\msun$ (2{,}000 haloes per 0.25-dex mass bin, ${\sim}22{,}900$ haloes total), drawn from the same 1\,\gpc\ filament catalogue. We use $M_{200\mathrm{c}}$ as the mass variable because $R_{200}$ is derived from it (equation~\ref{eq:r200}). We fit a linear relation in log space,
\begin{equation}
\log_{10} \kappa = A \log_{10} M_{200\mathrm{c}} + B,
\label{eq:kappa_mass}
\end{equation}
where $M_{200\mathrm{c}}$ is in units of $h^{-1}\,\msun$. The data are binned into 0.25-dex mass bins and we fit only bins with mean $\kappa > 2$, since $\kappa \le 2$ indicates a halo at the endpoint of a filament rather than a genuine node in the cosmic web. Error bars show bootstrap $1\sigma$ confidence intervals on the mean (1{,}000 resamples per bin), and the fit is weighted by the inverse bootstrap variance. The shaded band in Figure~\ref{fig:connectivity_mass} shows the interquartile (25th--75th percentile) range of the individual per-halo $\kappa$ values at the fiducial threshold, indicating the halo-to-halo scatter about the mean relation; we summarise the scatter non-parametrically because $\kappa$ is integer-valued and its per-bin distribution is skewed.

The power-law relation holds over nearly two decades in halo mass above the $\kappa > 2$ threshold (Figure~\ref{fig:connectivity_mass}). At our primary aperture of $1.5\,R_{200}$, the best-fitting slopes are $A = 0.273 \pm 0.005$, $0.303 \pm 0.005$, and $0.315 \pm 0.009$ for $n_\sigma = 6.0$, 6.5, and 7.0, respectively (Table~\ref{tab:extended_connectivity_20p}). The slopes are broadly stable across persistence thresholds, with $A \approx 0.273$--$0.315$. At a fixed aperture of 1\,\mpch, the slopes are shallower ($A = 0.132$--$0.158$), reflecting the mass dependence of $R_{200}$: a fixed physical aperture probes relatively deeper into the filament network of low-mass haloes. For the cluster-only subsample, Table~\ref{tab:connectivity_20p} reports slopes measured at $1.5\,R_{200}$ that are consistent with the extended sample within the uncertainties.

\begin{table}
\centering
\caption{Mass--connectivity slopes from the extended halo sample ($\sim$22,889 haloes, $10^{12}$--$10^{15.5}\,h^{-1}\,\msun$). Column $A$ is the slope of $\log\kappa = A\log M_{200\mathrm{c}}+B$ fitted to bins with mean $\kappa > 2$, weighted by bootstrap uncertainty on the mean (1{,}000 resamples per bin).}
\label{tab:extended_connectivity_20p}
\begin{tabular}{lcc}
\hline\hline
$n_\sigma$ & $A$ ($1.5\,R_{200}$) & $A$ (1 Mpc/$h$) \\
\hline
6.0 & $0.273\pm0.005$ & $0.132\pm0.004$ \\
6.5 & $0.303\pm0.005$ & $0.154\pm0.005$ \\
7.0 & $0.315\pm0.009$ & $0.158\pm0.009$ \\
\hline\hline
\end{tabular}
\end{table}

\begin{table}
\centering
\caption{Connectivity of the 324 most massive clusters in MDPL2, measured at $1.5\,R_{200}$, using postprocessed filaments from the 20-particle mass cut run (length filter $>1$ Mpc/$h$, merging, smoothing). Column $A$ gives the slope of the mass--connectivity relation $\log\kappa = A\,\log M_{200\mathrm{c}} + B$.}
\label{tab:connectivity_20p}
\begin{tabular}{lccccl}
\hline\hline
$n_\sigma$ & $\langle\kappa\rangle$ & Median & Min & Max & Slope $A$ \\
\hline
6.0 & 6.0 & 6 & 1 & 12 & $0.34 \pm 0.07$ \\
6.5 & 5.4 & 5 & 1 & 11 & $0.30 \pm 0.07$ \\
7.0 & 4.7 & 5 & 1 & 10 & $0.30 \pm 0.08$ \\
\hline\hline
\end{tabular}
\end{table}

\subsubsection{Comparison with literature}

A direct slope comparison with \citet{Codis2018} is not straightforward because they use a different parameterisation and a different connectivity definition. Their formula $\kappa \approx \frac{10}{3}\log_{10}(M/10^{11}\,h^{-1}\,\msun)$ expresses $\kappa$ as a linear function of $\log M$ (not a power law), and their connectivity is defined topologically as the number of saddle-point connections at the peak-patch boundary, a global, aperture-free measure. Applied to cluster masses ($\log M \sim 14$--$15$), this formula predicts $\kappa \approx 10$--$13$, substantially higher than our shell-crossing values of $\langle\kappa\rangle = 4.7$--$6.0$. The offset is expected: the shell-crossing definition counts only filaments that physically pierce the $1.5\,R_{200}$ aperture, whereas the topological definition counts all skeleton connections regardless of projected distance, inevitably yielding higher values.

\citet{Santoni2024} measured connectivity from \disperse applied to the gridded gas distribution from the GADGET-X hydrodynamical zoom-in resimulations of the same 324 Three Hundred clusters, finding $A = 0.308 \pm 0.019$ at $R_{200}$ (with $A = 0.295$--$0.324$ across apertures $R_{500}$, $R_{200}$, and $R_\mathrm{vir}$). Our extended sample slope at $1.5\,R_{200}$ is $A \approx 0.273$--$0.315$ (Table~\ref{tab:extended_connectivity_20p}), consistent with the gas-based result at $R_{200}$. \citet{Bermejo2024} characterised topological properties of the cosmic web using persistent homology (Betti curves), finding a slope $a = 0.33$ for filamentary structure, also consistent with our value. The close agreement between our dark matter halo-traced filaments and both the gas-traced filaments of \citet{Santoni2024} and the topological analysis of \citet{Bermejo2024}, despite very different tracers and methodologies, demonstrates that the mass-connectivity scaling at $A \approx 0.273$--$0.315$ is a robust feature of the cosmic web topology.

In terms of absolute values, \citet{2010Aragon} found $\kappa \sim 2$--$5$ for haloes in the mass range $10^{14}$--$10^{15}\,h^{-1}\,\msun$ using the Multiscale Morphology Filter. \citet{Malavasi2020} measured $\kappa \sim 2$--$3$ (median 2.5) for the Coma cluster using \disperse on SDSS observational data, while \citet{DarraghFord2019} confirmed the positive mass-connectivity correlation in the COSMOS survey. \citet{Malavasi2023} studied filaments connected to the Coma cluster using DisPerSE applied to both SDSS observations and constrained $N$-body simulations, finding $\kappa = 2.5$ (median, observed) and $\kappa = 3$--$4$ (simulated, depending on DisPerSE parameters), measured at $1.5\,r_\mathrm{vir}$ (${\approx}1.83\,R_{200}$ for typical overdensity parameters at $z=0$, a larger aperture than our $1.5\,R_{200}$); they further showed that dark-matter infall toward the cluster is significantly more collimated near filament spines than in the field. \citet{Kuchner2020} measured the properties of filaments in the outskirts of the same 324 Three Hundred clusters used here, using DisPerSE on galaxy mock catalogues from the GADGET-X zoom-in resimulations, finding filaments extending to ${\sim}5\,R_{200}$ around each cluster. Our mean values ($\langle \kappa \rangle = 4.7$--$6.0$ depending on the persistence threshold) are higher than the $\kappa \sim 2.5$--$4$ reported for Coma \citep{Malavasi2020, Malavasi2023}. This offset should not be attributed to a mass difference: our sample spans $M_\mathrm{vir} = 8 \times 10^{14}$--$3.5 \times 10^{15}\,h^{-1}\,\msun$ with a median of ${\approx}10^{15}\,h^{-1}\,\msun$, comparable to the Coma cluster ($M_{200} \approx 9 \times 10^{14}\,h^{-1}\,\msun$; \citealt{Malavasi2023}), and 63 per cent of our clusters have $M_\mathrm{vir}$ below the equivalent Coma virial mass. The discrepancy is instead driven by methodology: our filaments are detected from a simulation halo catalogue at the mean halo number density of the full 1\,\gpc\ box, whereas \citet{Malavasi2020} and \citet{Malavasi2023} use spectroscopic galaxy catalogues with a different tracer density and a different DisPerSE persistence threshold. Note that \citet{Malavasi2023} measure at $1.5\,r_\mathrm{vir} \approx 1.83\,R_{200}$, a larger aperture than our $1.5\,R_{200}$; the fact that our $\kappa$ values exceed theirs despite the smaller aperture rules out aperture choice as a driver of this offset. This highlights the importance of specifying the persistence threshold, tracer type, tracer number density, and aperture definition when comparing connectivity measurements across different analyses.

More recently, \citet{GalarragaEspinosa2024} measured connectivity from \disperse applied to the MillenniumTNG halo catalogue in a $(500\,h^{-1}\,\mathrm{Mpc})^3$ box, extending the analysis to multiple redshifts ($z = 0$--$2$) and across a broad mass range. Their slopes are consistent with our values, reinforcing the robustness of the mass--connectivity scaling across independent simulations, tracers, and cosmic epochs. On the observational side, \citet{EuclidGouin2025} presented the first connectivity measurement from the Euclid Quick Data Release (Q1), analysing ${\sim}220$ clusters at $0.2 < z < 0.7$ over 63\,deg$^2$. They recover the positive mass--connectivity relation independently of the filament detection algorithm, demonstrating that the scaling predicted by simulations is now accessible to wide-field photometric surveys. In a companion study, \citet{EuclidLaigle2025} used the same Euclid Q1 data to show that massive galaxies preferentially align their major axes along cosmic filaments, providing further evidence that the filamentary network shapes galaxy properties.

The broad stability of the slope $A \approx 0.273$--$0.315$ across persistence thresholds, despite the factor of $\sim$1.3 variation in mean $\kappa$, implies that increasing $n_\sigma$ removes filaments approximately uniformly across the mass range. This is consistent with the expectation that persistence reflects the local density contrast along the filament rather than the mass of the nodes it connects.

We note several limitations of the connectivity measurement. First, the absolute value of $\kappa$ depends on the persistence threshold, which has no unique physical prescription; different choices of $n_\sigma$ yield connectivity values that differ by up to a factor of $\sim$1.3, as shown in Table~\ref{tab:connectivity_20p}. Second, the shell-crossing definition of connectivity counts any filament arc that crosses the aperture, including tangential arcs that may not genuinely feed the cluster; alternative definitions based on the persistence skeleton topology may yield different results. Third, the extended mass-connectivity sample is drawn from a single simulation at $z = 0$; measuring the redshift evolution of the relation would strengthen the conclusions and is the subject of ongoing investigation.

\section{Discussion}
\label{sec:discussion}

\subsection{Systematic uncertainties of the frozen core method}

The frozen core pipeline achieves 99.6 per cent total length recovery and 94.7 per cent individual filament matching compared to a monolithic reference run (Section~\ref{sec:frozen_core}). Critical point recovery is even higher: 100 per cent of density maxima (peaks) and minima (void centres) are preserved, along with 99.7 per cent of 2-saddles, with zero spurious critical points introduced (Section~\ref{sec:frozen_core}). Since most downstream analyses use critical points as their starting point, this near-perfect recovery at the node level is the more relevant accuracy metric. The ${\sim}$5 per cent of unmatched filaments are predominantly short, low-significance structures (mean length 4.2\,\mpch). Short unmatched filaments are distributed throughout the tile, arising from tetrahedra whose circumspheres extend beyond the frozen core boundary at any location; only the long unmatched filaments are preferentially concentrated near tile boundaries, where missing context disrupts extended structures (Fig.~\ref{fig:tiling_unmatched}). This systematic is conservative: the method may miss some short filaments but does not introduce spurious structures into the catalogue. This stands in contrast to the no padding case, where the boundary removes 13.7 per cent of genuine filaments and also creates spurious structures from artificial critical points at the tessellation edge (Figure~\ref{fig:boundary_recovery}d). The padding and circumsphere filtering in the frozen core method eliminate both failure modes.

For statistical studies of the cosmic web filament length distributions, connectivity measurements, network topology this level of fidelity is sufficient. The $<$1 per cent systematic uncertainty in total filament length is smaller than the variation introduced by the choice of persistence threshold (Section~\ref{sec:nsigma}), which changes the total filament length by a factor of 2.3 across the range $n_\sigma = 6.0$--$7.0$. The method is less well suited to studies requiring exact filament configurations around individual objects; for such applications, the object of interest should be placed at the centre of a tile, far from boundaries, or a monolithic run on a smaller extraction region should be used.

\subsection{Comparison with naive sub volume approaches}

It is important to distinguish the frozen core method from the naive approach of running \disperse independently on smaller sub volumes without padding or tessellation filtering. As discussed in Section~\ref{sec:disperse_bottleneck}, smaller sub volumes produce different Delaunay tessellations with different network topologies, leading to different persistence pairs and different filaments. The controlled experiment in Section~\ref{sec:padding} quantifies this directly: running \disperse on a 300\,\mpch\ sub volume without padding removes 13.7 per cent of filaments and creates spurious structures near the boundaries, with the effect strongly length dependent longer filaments that extend across the boundary are preferentially affected, with recovery falling to ${\sim}$13 per cent for $L > 20$\,\mpch\ within 10\,\mpch\ of the edge (Figure~\ref{fig:boundary_recovery}b). The frozen core method avoids this problem entirely: by computing the tessellation on padded tiles and filtering with the circumsphere criterion, the merged complex is equivalent to the global Delaunay tessellation (to within the 0.37 per cent of tetrahedra whose circumspheres extend beyond the padding). The persistence pairs and resulting filaments are therefore statistically consistent with those that would be obtained from a monolithic run.

\subsection{Comparison with other filament detection methods}

The frozen-core method is specific to \disperse and other topological filament finders that require Delaunay tessellation as input. Other cosmic web classification algorithms face different computational constraints and have adopted different strategies for large-volume analysis. We focus on NEXUS+ and the Bisous model as representative alternatives because they illustrate the two main classes of non-topological cosmic web finders -- grid-based density field classification and stochastic geometric modelling -- and both have been widely applied to large-volume simulations.

Grid based methods such as NEXUS+ \citep{Cautun2013, Cautun2014} operate on density fields smoothed onto regular grids, with memory requirements scaling as the number of grid cells rather than the number of particles. These methods can readily analyse large volumes by choosing an appropriate grid resolution, but the fixed grid spacing limits the dynamic range of structures that can be resolved in a single pass and introduces preferred cardinal directions in the gradient tracing. \citet{Libeskind2018} compared twelve different cosmic web classification codes on the same simulation and found a factor of ${\sim}30$ scatter in the median dark matter overdensity assigned to filaments, reflecting the different definitions and scales probed by each method.

The Bisous model \citep{Tempel2014} uses a stochastic marked point process to identify filament spines from galaxy positions. Its computational cost scales with the number of galaxies and the number of Monte Carlo iterations, but it does not require global tessellation and can be applied to arbitrarily large volumes by processing overlapping subregions. However, the stochastic nature of the algorithm means that different realisations yield slightly different filament networks, introducing a source of variance that is absent in deterministic methods like \disperse.

The frozen core method preserves the deterministic, parameter free identification of \disperse while enabling its application to volumes that would otherwise be inaccessible. This is particularly important for studies that require the topological properties of the filament network such as connectivity measurements where the exact pairing of critical points determines the result.

\subsection{Context within current large-scale simulations}

The gigaparsec scale filament catalogue presented here complements recent efforts to characterise the cosmic web in large volumes. \citet{Cohn2022DisperseClusters} applied \disperse to a dark matter simulation to study cluster--node matching, finding that ${\sim}$3/4 of clusters have a \disperse node counterpart for smoothing $\leq 2.5$\,\mpch. \citet{Zhang2024FilamentStatistics} characterised filament length functions and radial density profiles in the ELUCID simulation using the Hessian-based COWS method. Both studies operate on volumes smaller than 1\,\gpch and use either different cosmic web finders or do not address the tiling problem that motivates the present work. \citet{GalarragaEspinosa2024} applied \disperse to the MillenniumTNG simulation \citep{Pakmor2023}, a $(500\,h^{-1}\,\mathrm{Mpc})^3$ box, identifying ${\sim}3300$ filaments at $z = 0$ with a mean length of ${\sim}12$\,Mpc. Our MDPL2 catalogue, at eight times the volume, contains $5 \times 10^5$--$2.3 \times 10^6$ filaments (depending on the persistence threshold), enabling statistical studies with substantially larger samples and access to rarer structures.

The frozen core method opens the possibility of applying \disperse to the next generation of cosmological simulations. The Euclid Flagship mock catalogue contains 3.4 billion galaxies in a volume of $(3780\,h^{-1}\,\mathrm{Mpc})^3$. With the tiling strategy demonstrated here, this volume could be processed in ${\sim}50{,}000$ tiles with the same per tile memory budget, requiring only proportionally more compute time. Similarly, halo catalogues from the FLAMINGO simulation suite, which span volumes up to $(2800\,h^{-1}\,\mathrm{Mpc})^3$, are now accessible to topologically rigorous filament analysis.

\subsection{Scalability and practical considerations}

The peak memory per tile depends on the local particle density and the tile plus padding volume, not on the total simulation size. The method therefore scales to arbitrarily large simulations by increasing the number of tiles while keeping the memory per tile constant. For the MDPL2 catalogue (92 million haloes in 1\,\gpc; Section~\ref{sec:mdpl2}), the 1000-tile configuration requires ${\sim}5$--$10$\,GB per tile. For a hypothetical 2\,\gpc\ box at similar particle density, the same tile size would require 8000 tiles with identical per tile memory, and the computation would remain embarrassingly parallel. The total runtime scales linearly with the number of tiles and can be reduced proportionally with additional compute nodes.

The choice of padding width $P$ involves a trade off between computational cost and topology preservation. Larger padding captures a higher fraction of tetrahedra whose circumspheres would otherwise extend beyond the tile boundary, but increases the volume (and hence memory) per tile as $(L_{\rm tile} + 2P)^3$ (equation~\ref{eq:volume_overhead}). As shown in Figure~\ref{fig:circumradius_dist} (right panel), the coverage curve exhibits diminishing returns: $P = 60$\,\mpch\ captures 99.33 per cent of tetrahedra for the MDPL2 halo distribution, while increasing $P$ to 100\,\mpch\ adds only 0.32 per cent additional coverage at substantially greater memory cost. For particle level analysis of higher resolution simulations, where the mean inter particle separation is smaller, a smaller padding width may suffice; conversely, sparse tracer populations may require larger padding. The circumradius distribution should be characterised for each new application to determine the appropriate $P$.

\subsection{Limitations}

Several limitations of the present work should be noted. First, the method has been validated only for halo catalogues; application to full particle distributions, which have different spatial statistics, may require adjustment of the padding width and tile size. Second, although the simplified pipeline has only one empirical parameter ($\delta = 0.5$\,\mpch) that lies in a stable regime (Appendix~\ref{sec:tolerance}), this stable range was established for the MDPL2 halo density (${\sim}0.09$ haloes per $(\mathrm{Mpc}\,h^{-1})^3$) and may need to be re evaluated for simulations with very different particle densities. Third, the connectivity measurements are limited to a single redshift ($z = 0$); extending the analysis to higher redshifts would probe the redshift evolution of the mass-connectivity relation.

Finally, we emphasise that the frozen core method does not accelerate \disperse itself: the Morse Smale extraction and persistence computation still operate on the full merged tessellation. The method reduces peak memory during the Delaunay tessellation stage, which is the primary bottleneck for large catalogues, but the subsequent \disperse stages must still process the full simplicial complex. For the MDPL2 catalogue (92 million haloes), the MSE step requires memory proportional to the number of simplices in the merged tessellation, which remains substantial ($\mathcal{O}(10^2)$\,GB) and may be prohibitive for some computing environments.

\section{Conclusions}
\label{sec:conclusions}

We have presented a frozen core method that enables \disperse to analyse extremely large cosmological simulations while preserving the network topology required for persistence-based filament detection. Our main results are as follows.

\begin{enumerate}
    \item \textbf{The computational bottleneck.} The three dimensional Delaunay tessellation required by \disperse scales as ${\sim}5$--$10$\,GB per million input particles, making monolithic runs infeasible for catalogues exceeding ${\sim}10^7$ objects. Running \disperse on smaller sub volumes does not solve this problem, because different volumes produce different Delaunay tessellations and therefore different persistence pairs and filaments (Section~\ref{sec:disperse_bottleneck}).

    \item \textbf{The frozen core method.} We decompose the simulation volume into overlapping tiles, compute the Delaunay tessellation independently on each tile with bounded memory, and apply a circumsphere-based filter that retains only tetrahedra guaranteed to belong to the global tessellation (Section~\ref{sec:frozen_core}). A two-stage post-processing pipeline (spatial deduplication and boundary stitching) merges the tiled \disperse outputs with 99.6 per cent total length recovery, 100 per cent recovery of density maxima and minima, and 94.7 per cent individual filament matching compared to a monolithic reference run. Systematic pipeline simplification analysis (Table~\ref{tab:pipeline_simplification}) demonstrates that deduplication is essential (removes 10 per cent excess length from tile overlaps) and stitching improves topology without affecting total length. The simplified pipeline has a single empirical parameter ($\delta = 0.5$\,\mpch), which lies in a stable regime (Appendix~\ref{sec:parameter_sensitivity}).

    \item \textbf{Application to MDPL2.} We apply the method to the full $(1\,h^{-1}\,\mathrm{Gpc})^3$ MDPL2 simulation (92 million haloes after a 20-particle mass cut), producing filament catalogues at three persistence thresholds ($n_\sigma = 6.0$, 6.5, 7.0). The catalogues contain $5 \times 10^5$--$2.3 \times 10^6$ filaments with total lengths of $4.7 \times 10^6$--$1.1 \times 10^7$\,\mpch, providing a quantitative characterisation of the cosmic web network across the largest volume analysed to date with topologically rigorous methods (Section~\ref{sec:results}).

    \item \textbf{Mass-connectivity relation.} We measure the connectivity $\kappa$ for ${\sim}22{,}900$ haloes spanning $M_{200\mathrm{c}} = 10^{12}$--$10^{15.5}\,h^{-1}\,\msun$, from galaxy groups to the most massive clusters in MDPL2 (including the 324 parent haloes of The Three Hundred project). Connectivity is measured on the five-step post-processed filament catalogue (including merging and 10-iteration smoothing), consistent with the companion study of group accretion modes (Singh et al., in preparation). Fitting only bins with mean $\kappa > 2$ (excluding unphysical filament endpoints), the mass-connectivity relation follows a single power law with slope $A \approx 0.273$--$0.315$ at $1.5\,R_{200}$ across the three persistence thresholds, consistent with gas-based measurements from cluster zoom-in simulations ($A = 0.308 \pm 0.019$; \citealt{Santoni2024}) and with the topological bias slope of $a = 0.33$ from persistent homology \citep{Bermejo2024}. At a fixed aperture of $1\,h^{-1}\,\mathrm{Mpc}$, the slopes are shallower ($A = 0.132$--$0.158$), isolating the intrinsic mass dependence from the aperture scaling with $R_{200}$ (Section~\ref{sec:connectivity}).
\end{enumerate}

The frozen core method removes the memory bottleneck that has prevented topologically rigorous cosmic web analysis of the largest cosmological simulations. The method scales to extremely large volumes by increasing the number of tiles, and the computation is embarrassingly parallel. In a companion study (Singh et al., in preparation), we use this filament catalogue to classify galaxy groups by their proximity to cluster-connected filaments and study whether groups accreting along filaments exhibit distinct kinematic and structural properties compared to those infalling isotropically.

\section*{Acknowledgements}

AS, MEG, and FRP acknowledge financial support from the UK Science and Technology Facilities Council (STFC; grant ref ST/X000982/1).

The CosmoSim database used in this paper is a service by the Leibniz Institute for Astrophysics Potsdam (AIP). The MultiDark database was developed in cooperation with the Spanish MultiDark Consolider Project CSD2009 00064. The authors gratefully acknowledge the Gauss Centre for Supercomputing e.V. (www.gauss centre.eu) and the Partnership for Advanced Supercomputing in Europe (PRACE, www.prace ri.eu) for funding the MultiDark simulation project by providing computing time on the GCS Supercomputer SuperMUC at Leibniz Supercomputing Centre (LRZ, www.lrz.de). The Bolshoi simulations have been performed within the Bolshoi project of the University of California High Performance AstroComputing Center (UC HiPACC) and were run at the NASA Ames Research Center.

\section*{Data Availability}

The MDPL2 simulation is publicly available via CosmoSim (\texttt{https://www.cosmosim.org}).

%%%%%%%%%%%%%%%%%%%%%%%%%%%%%%%%%%%%%%%%%%%%%%%%%%

\bibliographystyle{mnras}
\bibliography{paper_refs.bib}

%%%%%%%%%%%%%%%%%%%%%%%%%%%%%%%%%%%%%%%%%%%%%%%%%%

\appendix

\section{Parameter Sensitivity Analysis}
\label{sec:parameter_sensitivity}

The frozen core pipeline described in Section~\ref{sec:frozen_core} has a single empirical parameter, the matching tolerance $\delta = 0.5$\,\mpch. Here we present the detailed sensitivity analyses that justify this choice and demonstrate that additional anti loop protections are unnecessary.

\subsection{Anti loop protections}
\label{sec:antiloop}

An earlier version of the pipeline included two additional safeguards against artificial loop formation during boundary stitching: (i)~a pre filter blocking connections where the resulting filament would form a closed loop (start$_i \approx$ end$_j$ within 10\,\mpch), and (ii)~a post filter rejecting stitched groups with length/extent ratio $> 2.5$. We tested all four on/off combinations of these protections (Table~\ref{tab:antiloop_necessity}) and found they have no material effect on the results: total length recovery, maximum filament length, and the number of super filaments are essentially identical with or without them. The pre filter blocks 954 potential connections but this changes only the filament count (by ${\sim}$930), not the aggregate statistics. The post filter is entirely redundant, rejecting no groups. We therefore omit both from the production pipeline, reducing the number of tuneable parameters from four to one.

\begin{table}
\centering
\caption{Effect of anti loop protections on stitching. Total length recovery and maximum filament length are identical across all configurations; neither protection is necessary for this dataset.}
\label{tab:antiloop_necessity}
\begin{tabular}{lcccc}
\hline\hline
Configuration & $N_\mathrm{fil}$ & Length rec.\ (\%) & Max $L$ (\mpch) & $N_\mathrm{super}$ \\
\hline
Both on   & 31,667 & 99.56 & 67.1 & 6 \\
Pre filter only & 31,667 & 99.56 & 67.1 & 6 \\
Post filter only & 31,155 & 99.56 & 67.1 & 6 \\
Both off  & 31,155 & 99.56 & 67.1 & 6 \\
\hline\hline
\end{tabular}
\end{table}

\subsection{Matching tolerance}
\label{sec:tolerance}

The tolerance $\delta$ controls both centre based deduplication (Stage~2) and endpoint based boundary stitching (Stage~3). To assess the sensitivity of the pipeline to this parameter, we vary $\delta$ independently for each stage while holding the other fixed at 0.5\,\mpch.

\begin{figure*}
\centering
\includegraphics[width=\textwidth]{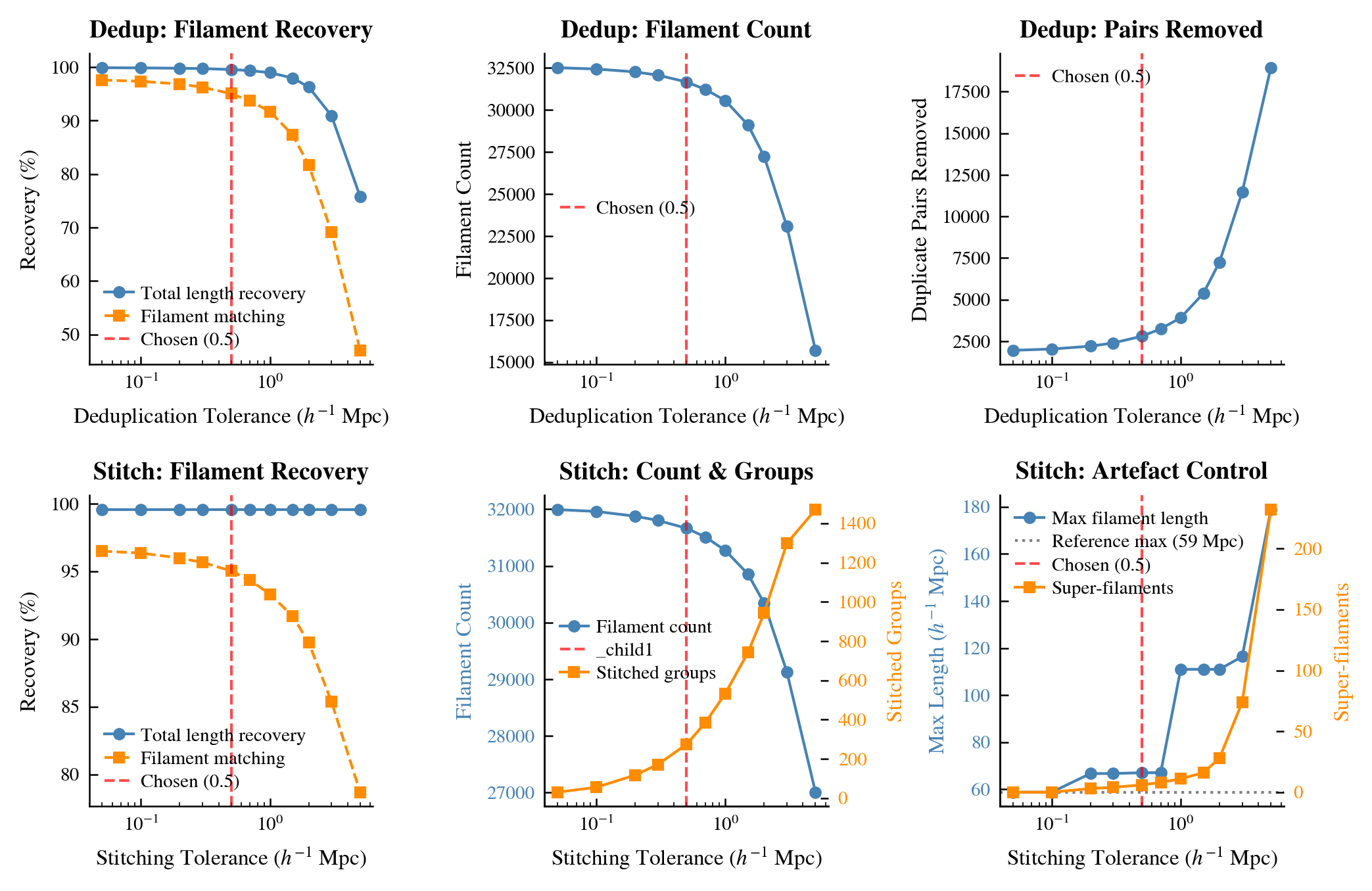}
\caption{\textbf{Sensitivity of the frozen core pipeline to the matching tolerance $\delta$.} \textbf{Top row:} Deduplication tolerance (stitching fixed at 0.5\,\mpch); recovery is stable for $\delta \leq 0.7$\,\mpch. \textbf{Bottom row:} Stitching tolerance (deduplication fixed at 0.5\,\mpch); above $\delta \sim 1$\,\mpch, over-stitching creates spurious super-filaments. Red dashed lines mark the chosen $\delta = 0.5$\,\mpch.}
\label{fig:tolerance_sensitivity}
\end{figure*}

Figure~\ref{fig:tolerance_sensitivity} and Table~\ref{tab:tolerance_sensitivity} show the results. For deduplication, length recovery exceeds 99.4 per cent for $\delta \leq 0.7$\,\mpch\ and degrades at larger values as distinct filaments are incorrectly merged (76 per cent recovery at $\delta = 5.0$\,\mpch). For stitching, the total length recovery is constant at 99.56 per cent regardless of $\delta$, but above $\delta = 1.0$\,\mpch\ over stitching creates spurious super filaments (232 at $\delta = 5.0$\,\mpch\ with maximum length 179\,\mpch). The chosen value $\delta = 0.5$\,\mpch\ lies well within the stable regime for both stages.

\begin{table}
\centering
\caption{Tolerance sensitivity analysis. \textbf{Top:} Deduplication tolerance (stitching fixed at 0.5\,\mpch). \textbf{Bottom:} Stitching tolerance (deduplication fixed at 0.5\,\mpch). The chosen value $\delta = 0.5$\,\mpch\ (bold) lies in the stable regime for both.}
\label{tab:tolerance_sensitivity}
\setlength{\tabcolsep}{4pt}
\begin{tabular*}{\columnwidth}{@{\extracolsep{\fill}}ccccc}
\hline\hline
\multicolumn{5}{c}{\textit{Deduplication tolerance}} \\
$\delta$ & $N_\mathrm{fil}$ & Len.\ rec. & Pairs rem. & Max $L$ \\
(\mpch) & & (\%) & & (\mpch) \\
\hline
0.1 & 32,431 & 99.86 & 2,036 & 67.1 \\
0.2 & 32,261 & 99.80 & 2,206 & 67.1 \\
0.5 & 31,664 & 99.56 & 2,803 & 67.1 \\
1.0 & 30,558 & 98.97 & 3,915 & 67.1 \\
2.0 & 27,239 & 96.27 & 7,260 & 67.1 \\
5.0 & 15,706 & 75.80 & 18,939 & 67.1 \\
\hline\hline
\multicolumn{5}{c}{\textit{Stitching tolerance}} \\
$\delta$ & $N_\mathrm{fil}$ & Stitched & $N_\mathrm{super}$ & Max $L$ \\
(\mpch) & & pairs & & (\mpch) \\
\hline
0.1 & 31,960 & 73 & 0 & 58.8 \\
0.2 & 31,877 & 156 & 3 & 66.7 \\
0.5 & 31,664 & 369 & 6 & 67.1 \\
1.0 & 31,272 & 761 & 11 & 111.0 \\
2.0 & 30,341 & 1,692 & 28 & 111.0 \\
5.0 & 27,005 & 7,352 & 232 & 178.8 \\
\hline\hline
\end{tabular*}
\end{table}

\subsection{Persistence threshold calibration via critical point-halo matching}
\label{sec:cp_halo_calibration}

The persistence threshold $n_\sigma$ is the primary free parameter in \disperse. To provide a physics-driven validation of our fiducial choice $n_\sigma = 6.5$, we adopt the critical point-halo matching method introduced by \citet{GalarragaEspinosa2020} and developed further by \citet{GalarragaEspinosa2024} and \citet{BoldriniLaigle2024}. The method compares the density-field maxima (CP$_{\rm max}$, critical points with $\texttt{cptype} = 3$) identified by \disperse against dark matter haloes from an independent catalogue, using two complementary metrics:

\begin{itemize}
    \item \textbf{Purity:} the fraction of CP$_{\rm max}$ whose nearest halo is within that halo's $R_{200}$. High purity indicates that \disperse is not placing spurious maxima in the density field.
    \item \textbf{Completeness:} the fraction of haloes whose nearest CP$_{\rm max}$ falls within the halo's $R_{200}$. High completeness indicates that \disperse is detecting the density peaks associated with real massive structures.
\end{itemize}

\noindent The optimal threshold is where the ratio $R = \text{purity}/\text{completeness} \approx 1$, indicating a balance between spurious detections and missed structures \citep{BoldriniLaigle2024}.

\begin{figure*}
\centering
\includegraphics[width=\textwidth]{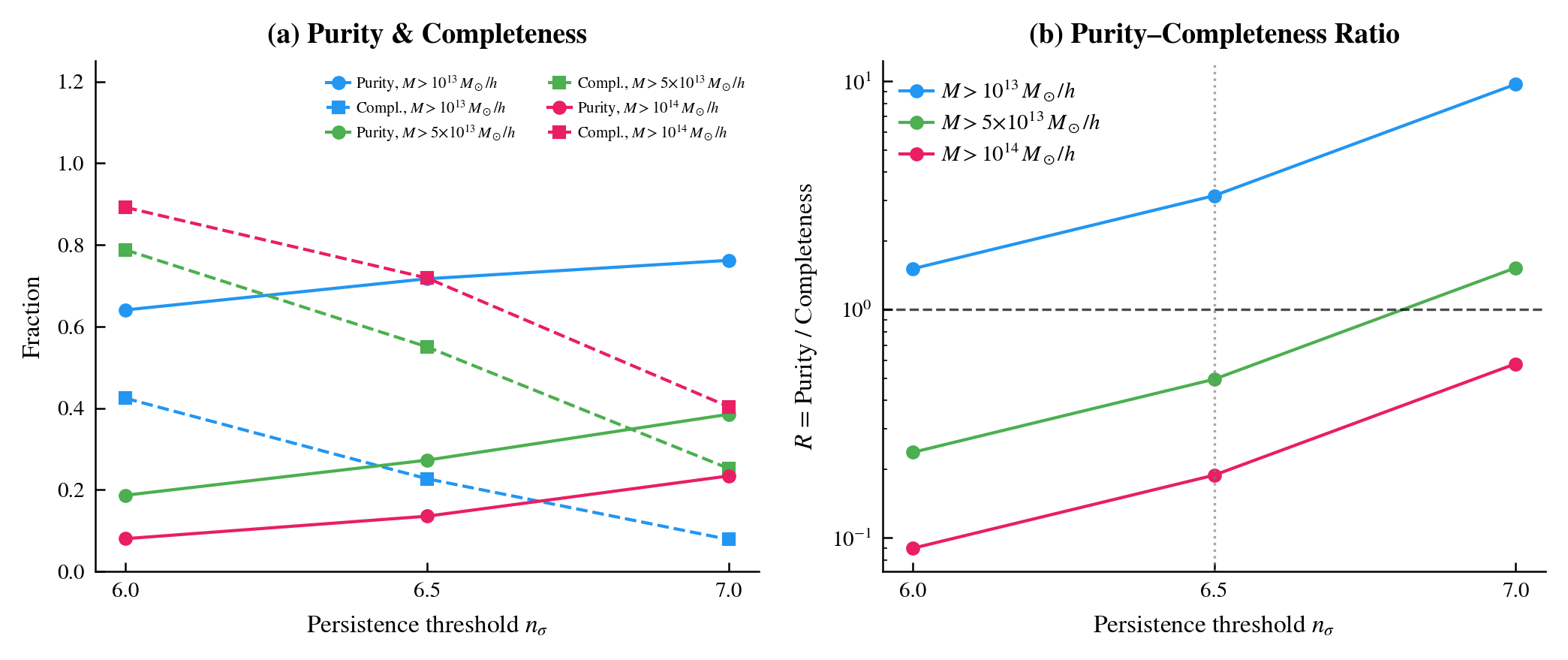}
\caption{Critical point-halo matching calibration of the persistence threshold. \textbf{(a)}~Purity (solid circles) and completeness (dashed squares) as a function of $n_\sigma$ for three halo mass cuts: $M_{\rm vir} > 10^{13}\,h^{-1}\,\msun$ (blue), $M_{\rm vir} > 5 \times 10^{13}\,h^{-1}\,\msun$ (green), and $M_{\rm vir} > 10^{14}\,h^{-1}\,\msun$ (pink). \textbf{(b)}~Ratio $R = \text{purity}/\text{completeness}$. The horizontal dashed line marks $R = 1$ (optimal balance). For the $M > 10^{13}$ sample, $R \approx 1$ at $n_\sigma \approx 6$--$6.5$, validating our fiducial choice of $n_\sigma = 6.5$. The higher mass cuts show $R$ crossing unity at larger $n_\sigma$; this reflects the artificially low purity when CP$_{\rm max}$ at lower-mass (but genuine) nodes are counted as impure, rather than indicating a preference for higher thresholds (see text).}
\label{fig:cp_halo_calibration}
\end{figure*}

\subsubsection{Choice of mass threshold}

The mass threshold for the halo sample must be chosen so that the selected haloes are genuine topological maxima of the density field, nodes of the cosmic web, rather than objects residing along filaments. To determine this empirically, we compute the median distance from haloes to their nearest CP$_{\rm max}$ as a function of halo mass at each persistence threshold. Table~\ref{tab:cp_mass_threshold} shows the result: at $n_\sigma = 6.0$, haloes above $M_{\rm vir} \sim 5 \times 10^{13}\,h^{-1}\,\msun$ have a median distance of effectively zero to the nearest CP$_{\rm max}$. At our fiducial $n_\sigma = 6.5$, this remains true for the most massive haloes ($M_{\rm vir} > 5 \times 10^{13}\,h^{-1}\,\msun$), but at $n_\sigma = 7.0$ the median distance increases to ${\sim}9$\,\mpch\ for the same mass cut, indicating that many genuine density maxima are no longer detected. This identifies $n_\sigma = 6.5$ as the threshold that preserves the correspondence between haloes and CP$_{\rm max}$ for massive structures, consistent with the cluster threshold adopted by \citet{GalarragaEspinosa2024} ($M_{200c} \geq 5 \times 10^{13}\,h^{-1}\,\msun$).

\begin{table}
\centering
\caption{Median distance from haloes to their nearest CP$_{\rm max}$ at each persistence threshold (20-particle mass cut). Above $M_{\rm vir} \sim 5 \times 10^{13}\,h^{-1}\,\msun$, haloes coincide with \disperse\ density maxima at $n_\sigma \leq 6.5$.}
\label{tab:cp_mass_threshold}
\begin{tabular}{lcccc}
\hline\hline
Mass cut & $N_{\rm halo}$ & $n_\sigma = 6.0$ & $n_\sigma = 6.5$ & $n_\sigma = 7.0$ \\
& & \multicolumn{3}{c}{Median distance (\mpch)} \\
\hline
$M > 10^{13}$ & 452,935 & 2.16 & 5.80 & 11.88 \\
$M > 5 \times 10^{13}$ & 71,121 & $<$0.01 & $<$0.01 & 9.15 \\
$M > 10^{14}$ & 27,029 & $<$0.01 & $<$0.01 & 5.85 \\
\hline\hline
\end{tabular}
\end{table}

\subsubsection{Results}

We compute purity and completeness at three persistence thresholds ($n_\sigma = 6.0$, 6.5, 7.0) using the CP$_{\rm max}$ from the full 1\,\gpc\ box and three halo mass cuts ($M_{\rm vir} > 10^{13}$, $5 \times 10^{13}$, and $10^{14}\,h^{-1}\,\msun$). Matching uses periodic nearest-neighbour queries (\texttt{cKDTree} with \texttt{boxsize}$= 1000$\,\mpch) and the variable $R_{200}$ radius of each halo (computed from $M_{200\mathrm{c}}$ via equation~\ref{eq:r200}).

Figure~\ref{fig:cp_halo_calibration} presents the results. Panel~(a) shows that purity increases monotonically with $n_\sigma$ (fewer spurious CP$_{\rm max}$ at higher thresholds), while completeness decreases (more haloes lack a nearby CP$_{\rm max}$). The behaviour is qualitatively consistent across all three mass cuts. Panel~(b) shows the ratio $R = \text{purity}/\text{completeness}$ on a logarithmic scale.

The $R = 1$ crossing point depends on the mass cut used. With the 20-particle mass cut, the ratio $R > 1$ at all three thresholds for $M > 10^{13}\,h^{-1}\,\msun$, reflecting the reduced number of spurious low-density CP$_{\rm max}$ compared to the full halo catalogue. For the higher mass cuts ($M > 5 \times 10^{13}$ and $M > 10^{14}\,h^{-1}\,\msun$), the crossing shifts to $n_\sigma \approx 6.5$--$7.0$. This shift does not imply that higher thresholds are more appropriate; rather, it reflects the different roles of the two metrics at different mass scales. When purity is measured against a high mass cut, CP$_{\rm max}$ that correspond to genuine but lower-mass nodes ($M_{\rm vir} \sim 10^{13}$--$5 \times 10^{13}\,h^{-1}\,\msun$) are counted as ``impure'' even though they are real density peaks. The purity measured against these restrictive cuts is therefore artificially depressed by the halo mass function, not by spurious detections.

The $M > 10^{13}\,h^{-1}\,\msun$ sample is the appropriate reference for the $R = 1$ \emph{calibration} because it includes all haloes massive enough to be genuine topological maxima of the density field. This mass threshold is distinct from the lower limit used for the \emph{connectivity} measurements in Section~\ref{sec:connectivity}, where we extend the sample down to $M_{200\mathrm{c}} = 10^{12}\,h^{-1}\,\msun$ to probe the mass--connectivity relation across group scales; those haloes need not be density maxima themselves to have filaments crossing their $R_{200}$ shell. At $n_\sigma = 6.5$, the number of CP$_{\rm max}$ (143{,}499) is comparable to the number of haloes above ${\sim}3 \times 10^{13}\,h^{-1}\,\msun$ (133{,}865), and the $M > 10^{13}$ sample provides the broadest baseline for testing both whether CP$_{\rm max}$ land on real structures (purity) and whether real nodes are being detected (completeness). This is consistent with the calibration approach of \citet{BoldriniLaigle2024}, who use $M > 10^{13}\,\msun$ for the ratio method.

The higher mass cuts serve a complementary purpose: they confirm that \disperse density maxima are \emph{physically meaningful}. The fact that haloes above $5 \times 10^{13}\,h^{-1}\,\msun$ have a median distance of effectively zero to the nearest CP$_{\rm max}$ (Table~\ref{tab:cp_mass_threshold}) demonstrates that massive clusters reliably coincide with \disperse nodes. This is a sanity check on the physical correspondence, independent of the threshold selection.

This analysis confirms that $n_\sigma = 6.5$ is an appropriate persistence threshold for the MDPL2 halo catalogue: it balances purity and completeness for the $M > 10^{13}\,h^{-1}\,\msun$ population, while preserving the filamentary bridges between groups and clusters that constitute the backbone of the cosmic web. Adopting a higher threshold (e.g.\ $n_\sigma = 7$) would improve purity at the cost of losing ${\sim}$58 per cent of all filaments (Table~\ref{tab:nsigma_comparison}), removing genuine inter-group structures rather than noise.

% Don't change these lines
% \bsp	% typesetting comment
\label{lastpage}
\end{document}